\documentstyle[aps,prb,graphicx,feynmp,amsbsy,amssymb,eqsecnum,multicol]{revtex}

\newcommand{\Sp}{\mathop{\rm Sp}\nolimits}
\newcommand{\etwocol}{\end{multicols}\widetext\begin{multicols}{2}\narrowtext%
            \noindent\hrulefill\unitlength1pt\line(0,1){6}\end{multicols}\widetext}
\newcommand{\btwocol}{\widetext\begin{multicols}{2}\noindent\hfill\narrowtext\par
            \noindent\unitlength1pt\line(0,-1){6}\hrulefill\end{multicols}\widetext
            \begin{multicols}{2}}
\tighten
\begin{document}
\preprint{ICMP--99--20E}
\draft
\title{Strong-coupling approach for strongly correlated electron systems}
\author{Andrij M. Shvaika}
\address{Institute for Condensed Matter Physics,
National Academy of Sciences of Ukraine, \\
1~Svientsitskii Str., UA--79011 Lviv, Ukraine}

\maketitle

\begin{abstract}
A perturbation theory scheme in terms of electron hopping,
which is based on the Wick theorem for Hubbard operators, is developed.
Diagrammatic series contain single-site vertices connected by hopping lines
and it is shown that for each vertex the problem splits
into the subspaces with ``vacuum states'' determined by the diagonal
Hubbard operators and only excitations around these vacuum states are
allowed. The rules to construct diagrams are proposed.
In the limit of infinite spatial dimensions the total auxiliary
single-site problem exactly splits into subspaces that
allows to build an analytical thermodynamically consistent approach
for a Hubbard model. Some analytical results are given for the simple
approximations when the two-pole (alloy-analogy solution) and four-pole
(Hartree-Fock approximation) structure for Green's function is
obtained. Two poles describe contribution from the Fermi-liquid
component, which is dominant for small electron and hole
concentrations (``overdoped case'' of high-$T_c$'s),
whereas other two describe contribution
from the non-Fermi liquid and are dominant close to half-filling
(``underdoped case'').
\end{abstract}

\begin{multicols}{2}
\narrowtext

\section{Introduction}

Many unconventional properties (e.g., metal-insulator transition, electronic
(anti)fer\-ro\-mag\-ne\-tism) of the nar\-row-band systems (transition metals and
their compounds, some organic systems, high-$T_c$ superconductors,
etc.)
can be explained only by the proper treatment of the strong local electron
correlations. The simplest models allowing for the electron correlations are
a single-band Hubbard model with on-site repulsion $U$ and hopping energy
$t$ and its strong-coupling limit ($U\gg t$): $t-J$ model. Recent studies of
the Hubbard-type models connected mainly with the theory of
high-$T_c$
superconductivity and performed in the weak- $(U\le 4t)$ and strong- $(U\gg t)$
coupling limits, elucidate some important features of these
models.\cite{DagottoReview} But still a lot
of problems remains, especially for the $U\gg t$ case where there are no
rigorous approaches.

Such approaches can be built using systematic
perturbation expansion in terms of the electron hopping\cite{IzyumovCMP} using
diagrammatic technique for Hubbard operators.\cite{Slobodyan,IzyumovBook}
One of them was proposed for the Hubbard ($U=\infty$ limit)
and $t-J$ models.\cite{IzyumovLetfulov,IzyumovPRB} The lack of such
approach is connected with the concept of a ``hierarchy'' system for Hubbard
operators when the form of the diagrammatic series and final results
strongly depend on the system of the pairing priority for Hubbard operators.
On the other hand it is difficult to generalize it on the case of the arbitrary
$U$.

In the last decade the essential achievements in the theory of the strongly
correlated electron systems are connected with the development of the
dynamical mean-field theory (DMFT) proposed by Metzner and
Vollhardt\cite{MetznerVollhardt} for the
Hubbard model (see also Ref.~\onlinecite{DMFTreview} and references therein).
DMFT is a
nonperturbative scheme that allows to project the Hubbard model on the single
impurity Anderson model and is exact in the limit of infinite space
dimensions $(d=\infty)$. There are no restrictions on the $U$ value within
this theory and it turns out to be useful for intermediate coupling ($U\sim
t$) for which it ensures the correct description of the metal-insulator
phase transition and determines the region of the Fermi-liquid behavior of
the electron subsystem. Moreover, some class of the binary-alloy-type models
(e.g., the Falicov-Kimball model) can be studied almost
analytically within DMFT.\cite{BrandtMielsch}
But in the case of the Hubbard model, the treatment of the effective single
impurity Anderson model is very complicated and mainly computer simulations
[exact diagonalization of the finite-sized systems or quantum Monte Carlo (QMC)]
are used, which calls for the development of the analytical approaches.

The first analytical approximation proposed for the Hubbard model was
a simple Hubbard-I approximation\cite{H-I} (see Ref.~\onlinecite{Dorneich}
for its possible improvement) which is correct in the atomic
($t=0$) and band ($U=0$) limits but is inconsistent in the
intermediate cases and cannot describe metal-insulator
transition. Hubbard's alloy-analogy solution\cite{H-III}
(so-called Hubbard-III approximation) incorporates into the theory
an electron scattering on the charge and spin fluctuations that
allows us to give qualitative description of the changes of the
density-of-state at the metal-insulator transition point.
Hubbard-I and Hubbard-III approximations introduces two types of
particles (electrons moving between empty sites and electrons moving
between sites occupied by electrons of opposite spin) with the
different energies that differ by $U$ and form two Hubbard bands.
Related schemes of the so-called two-pole
approximations,\cite{Roth,Nolting} which are justified
by the $t/U\ll 1$ perturbation theory expansions,\cite{H_L}
are also considered.
However, in
the recent QMC studies\cite{Grober,Pairault} it is clearly distinguished
four bands in the spectral functions rather than the two bands
predicted by the two-pole approximations. Such four-band structure is
reproduced by the strong-coupling expansion for the Hubbard
model\cite{Pairault} in the one-dimensional case. Within other
approaches let us mention non-crossing
approximation,\cite{Pruschke,Obermeier} Edwards-Hertz
approach,\cite{Edwards,Wermbter} iterative perturbation
theory,\cite{Kajueter,Wegner} alloy-analogy based
approaches,\cite{Herrmann,Potthoff} and linked cluster
expansions,\cite{Metzner,Janis}
which are reliable in certain limits and the construction of the
thermodynamically consistent theory still remains open.\cite{Gebhard}

The aim of this paper is to develop for Hubbard-type models a rigorous
perturbation theory scheme in terms of electron hopping that is based on
the Wick theorem for Hubbard operators\cite{Slobodyan,IzyumovBook}
and is valid for arbitrary value of
$U$ ($U<\infty$) and does not depend on the ``hierarchy'' system for $X$
operators. In the limit of infinite spatial dimensions, these analytical
schemes allow us to build a self-consistent Kadanoff-Baym type theory\cite{Baym}
for the Hubbard model and some analytical results are given for simple
approximations. The Falicov-Kimball model is also considered as an exactly
soluble limit of Hubbard model.

\section{Perturbation theory in terms of electron hopping}

We consider the lattice electronic system that can be described by the
following statistical operator:
\begin{eqnarray}
\hat{\rho} &=& e^{-\beta \hat{H}_{0}}\hat{\sigma}(\beta),
\label{eq1}\\
\hat{\sigma}(\beta) &=& T \!\exp \left\{\!-\!\int\limits_{0}^{\beta} \!\!d\tau
\!\int\limits_{0}^{\beta}\!\!d\tau' \sum\limits_{ij\sigma} t_{ij}^{\sigma}
(\tau-\tau') a_{i\sigma}^{\dag}(\tau) a_{j\sigma}(\tau') \right\},
\nonumber
\end{eqnarray}
where
\begin{equation}
\hat{H}_{0} = \sum\limits_{i} \hat{H}_{i}
\label{eq2}
\end{equation}
is a sum of the single-site contributions and for the Hubbard model we must put
\begin{eqnarray}
& H_{i}=Un_{i\uparrow} n_{i\downarrow} - \mu(n_{i\uparrow} +
n_{i\downarrow}) - h(n_{i\uparrow} - n_{i\downarrow}),
\nonumber\\
& t_{ij}^{\sigma} (\tau-\tau') = t_{ij} \delta (\tau-\tau').
\label{eq3}
\end{eqnarray}
In addition for the Falicov-Kimball model we must put
\begin{equation}
t_{ij}^{\sigma}(\tau-\tau') = \left\{\begin{array}{cl}
t_{ij} \delta(\tau-\tau') & \text{ for $\sigma=\uparrow$}\\
0 & \text{ for $\sigma=\downarrow$}
\end{array}
\right..
\label{eq4}
\end{equation}

It is supposed that we know eigenvalues and eigenstates of the zero-order
Hamiltonian (\ref{eq2}),
\begin{equation}
H_{i}|i,p\rangle = \lambda_{p} |i,p\rangle
\end{equation}
and one can introduce Hubbard operators
\begin{equation}
\hat{X}_{i}^{pq}=|i,p\rangle\langle i,q|
\label{eq5}
\end{equation}
in terms of which zero-order Hamiltonian is diagonal
\begin{equation}\label{Hdiag}
H_{0}=\sum_{i}\sum_{p} \lambda_{p}\hat{X}_{i}^{pp}.
\end{equation}

For the Hubbard model we have four states
$|i,p\rangle=|i,n_{i\uparrow},n_{i\downarrow}\rangle$:
$|i,0\rangle=|i,0,0\rangle$ (empty site),
$|i,2\rangle = |i,1,1\rangle$ (double occupied site),
$|i,\uparrow\rangle=|i,1,0\rangle$ and $|i,\downarrow\rangle=|i,0,1\rangle$
(sites with spin-up and spin-down electrons) with energies
\begin{equation}
\lambda_{0}=0, \; \lambda_{2}=U-2\mu, \; \lambda_{\downarrow} =
h-\mu, \; \lambda_{\uparrow} = -h-\mu.
\label{eq6}
\end{equation}
The connection between the electron operators and the Hubbard operators is the following:
\begin{equation}
n_{i\sigma} = X_{i}^{22} + X_{i}^{\sigma\sigma}; \; a_{i\sigma} =
X_i^{0\sigma} + \sigma X_i^{\bar\sigma 2}.
\label{eq7}
\end{equation}

Our aim is to calculate the grand canonical potential functional
\begin{eqnarray}
&&\Omega=-\frac{1}{\beta} \ln \Sp\hat{\rho} = \Omega_{0} - \frac{1}{\beta} \ln
\langle\hat{\sigma}(\beta)\rangle_{0}, \nonumber\\
&&\Omega_{0}=-\frac{1}{\beta} \ln \Sp e^{-\beta H_{0}},
\label{eq8}
\end{eqnarray}
single-electron Green functions
\begin{equation}
G_{ij\sigma}(\tau-\tau') =\langle T a_{i\sigma}^{\dag}(\tau) a_{j\sigma}(\tau')
\rangle = \frac{\delta \Omega}{\delta t_{ij}^{\sigma}(\tau-\tau')}
\label{eq9}
\end{equation}
and mean values
\begin{eqnarray}
n_{\sigma}=\frac{1}{N} \sum_{i} \langle n_{i\sigma} \rangle  =
-\frac{1}{N} \frac{d\Omega}{d\mu_{\sigma}},
\nonumber\\
n=n_{\uparrow}+n_{\downarrow}; \;
m=n_{\uparrow}-n_{\downarrow},
\label{eq10}
\end{eqnarray}
where $\mu_{\sigma}=\mu+\sigma h$ is a chemical potential for the
electrons with spin $\sigma$.
Here, $\langle\ldots\rangle=(1/Z) \Sp(\ldots \hat{\rho})$,
$Z=\Sp\hat{\rho}$, or in interacting representation
\begin{equation}
\langle\ldots\rangle = \frac{1}{\langle \hat{\sigma}(\beta)\rangle_{0}}
\langle\ldots\hat{\sigma}(\beta)\rangle_{0} =
\langle\ldots\hat{\sigma}(\beta) \rangle_{0c},
\label{eq11}
\end{equation}
where $\langle\ldots\rangle_{0}=(1/Z_0) \Sp (\ldots e^{-\beta H_{0}})$;
$Z_{0}=\Sp e^{-\beta H_{0}}$.

We expand the scattering matrix $\hat{\sigma}(\beta)$ in Eq.~(\ref{eq1}) into the series
in terms of electron hopping and for $\langle \sigma(\beta)\rangle_{0}$ we
obtain a series of terms that are products of the hopping integrals and
averages of the electron creation and annihilation operators or, using
Eq.~(\ref{eq7}), Hubbard operators that will be calculated with the use of
the corresponding Wick's theorem.

Wick's theorem for Hubbard operators was formulated in Ref.~\onlinecite{Slobodyan}
(see also Ref.~\onlinecite{IzyumovBook} and references therein).
For the Hubbard model we can define four
diagonal Hubbard operators $X^{pp}$ ($p=0,2,\downarrow,\uparrow$) which
are of bosonic type, four
annihilation $X^{0\downarrow}$, $X^{0\uparrow}$, $X^{\uparrow 2}$,
$X^{\downarrow 2}$ and four conjugated creation fermionic operators, and two
annihilation $X^{\downarrow\uparrow}$, $X^{02}$ and two conjugated creation
bosonic operators. The algebra of $\hat{X}$ operators is defined by the
multiplication rule
\begin{equation}
X_{i}^{rs} X_{i}^{pq} = \delta_{sp} X_{i}^{rq},
\label{eq12}
\end{equation}
the conserving condition
\begin{equation}
\sum_p X_i^{pp}=1,
\label{eq12p}
\end{equation}
and the commutation relations
\begin{equation}
[X_{i}^{rs},X_{j}^{pq}]_\pm = \delta_{ij} (\delta_{sp} X_{i}^{rq} \pm
\delta_{rq} X_{i}^{ps}),
\label{eq13}
\end{equation}
where one must use anticommutator when both operators are of the fermionic
type and commutator in all other cases. So, commutator or anticommutator of
two Hubbard operators is not a $c$ number but a new Hubbard operator. Then
the average of a $T$ products of $X$ operators can be evaluated by the
consecutive  pairing, while taking into account standard permutation rules for
bosonic and fermionic operators, of all off-diagonal Hubbard operators $X^{pq}$
according to the rule (Wick's theorem)
\begin{equation}
\stackrel{\unitlength=1em\line(0,-1){.3}\vector(-1,0){1.8}\line(-1,0){1.4}\line(0,-1){.3}\quad}
{X_{i}^{rs}(\tau_{1}) X_{0}^{pq}}(\tau)
=- \delta_{0i} g_{pq} (\tau-\tau_{1})
[X_{i}^{rs}(\tau_{1}), X_{i}^{pq} (\tau_{1})]_\pm
\label{eq14}
\end{equation}
until we get the product of the diagonal Hubbard operators only. Here we
introduce the zero-order Green's function
\begin{eqnarray}
g_{pq}(\tau-\tau_{1})&=&
\frac1{\beta}\sum_n g_{pq}(\omega_n) e^{i\omega_n(\tau-\tau_1)}
\label{eq15}\\
&=&e^{(\tau-\tau_{1})\lambda_{pq}}
\left\{\begin{array}{ll}
\pm n_\pm(\lambda_{pq}) & \quad\tau > \tau_{1} \\
\pm n_\pm(\lambda_{pq})-1 & \quad\tau < \tau_{1}
\end{array} \right.,
\nonumber
\end{eqnarray}
where $\lambda_{pq}=\lambda_{p}-\lambda_{q}$ and $n_\pm(\lambda) =
\frac{1}{e^{\beta\lambda}\pm1}$, and its Fourier  transform is equal
\begin{equation}
g_{pq}(\omega_{n}) = \frac{1}{i\omega_{n}-\lambda_{pq}}.
\label{eq16}
\end{equation}

In particular, for the Hubbard model one can introduce the following
pairings:
\begin{eqnarray}
\stackrel{\unitlength=1em\line(0,-1){.3}\vector(-1,0){1.6}\line(-1,0){1.2}\line(0,-1){.3}\quad}
{a_{i\sigma}(\tau_1)a_{j\sigma}^{\dag}}\!\!(\tau)&=&-\delta_{ij}\big\{
g_{\sigma0}(\tau-\tau_1)(X_i^{00}(\tau_1)+X_i^{\sigma\sigma}(\tau_1))
\nonumber\\
&&\quad+g_{2\bar\sigma}(\tau-\tau_1)(X_i^{22}(\tau_1)+X_i^{\bar\sigma\bar\sigma}(\tau_1))\big\},
\nonumber\\
\stackrel{\unitlength=1em\line(0,-1){.3}\vector(-1,0){1.6}\line(-1,0){1.2}\line(0,-1){.3}\quad}
{a_{i\bar\sigma}(\tau_1)a_{j\sigma}^{\dag}}\!\!(\tau)&=&-\delta_{ij}
f_{\sigma}(\tau-\tau_1)X_i^{\sigma\bar\sigma}(\tau_1),
\nonumber\\
\stackrel{\unitlength=1em\line(0,-1){.3}\vector(-1,0){1.6}\line(-1,0){1.2}\line(0,-1){.3}\quad}
{a_{i\bar\sigma}^{\dag}(\tau_1)a_{j\sigma}^{\dag}}\!\!(\tau)&=&\delta_{ij}
f_{\sigma}(\tau-\tau_1)\cdot\sigma\cdot X_i^{20}(\tau_1),
\label{eq16p}\\
\stackrel{\unitlength=1em\line(0,-1){.3}\vector(-1,0){1.7}\line(-1,0){1.3}\line(0,-1){.3}\quad}
{a_{i\bar\sigma}^{\dag}(\tau_1)X_j^{\sigma\bar\sigma}}\!\!(\tau)&=&\delta_{ij}
g_{\sigma\bar\sigma}(\tau-\tau_1)a_{i\sigma}^{\dag}(\tau_1),
\nonumber\\
\stackrel{\unitlength=1em\line(0,-1){.3}\vector(-1,0){1.7}\line(-1,0){1.3}\line(0,-1){.3}\quad}
{a_{i\sigma}(\tau_1)X_j^{\sigma\bar\sigma}}\!\!(\tau)&=&-\delta_{ij}
g_{\sigma\bar\sigma}(\tau-\tau_1)a_{i\bar\sigma}(\tau_1),
\nonumber\\
\stackrel{\unitlength=1em\line(0,-1){.3}\vector(-1,0){1.7}\line(-1,0){1.3}\line(0,-1){.3}\quad}
{a_{i\sigma}(\tau_1)X_j^{20}}\!\!(\tau)&=&-\delta_{ij}
g_{20}(\tau-\tau_1)\cdot\sigma\cdot a_{i\bar\sigma}^{\dag}(\tau_1),
\nonumber
\end{eqnarray}
where
\begin{eqnarray}
f_{\sigma}(\omega_n)&\equiv&g_{\sigma0}(\omega_n)-g_{2\bar\sigma}(\omega_n)
\\
&=&-U g_{\sigma0}(\omega_n) g_{2\bar\sigma}(\omega_n).
\nonumber
\end{eqnarray}

Applying such pairing procedure to the expansion of
$\langle\hat{\sigma}(\beta)\rangle_{0}$ we get the following diagrammatic
representation:
\etwocol
\widetext
\begin{eqnarray}
\left\langle\hat\sigma(\beta)\right\rangle_0&=&\left\langle
\exp\Biggl\{
\right.
-\;
\unitlength=0.1em
\parbox{47\unitlength}{
    \begin{fmffile}{loop1c}
    \begin{fmfgraph}(40,25)\fmfkeep{loopa}
        \fmfpen{thin}\fmfleft{el}\fmfright{er}
        \fmf{photon,right=.7}{er,el}
        \fmf{fermion}{er,el}\fmf{phantom}{er,c1,c2,el}
        \fmf{plain,left,width=thin,tension=.2}{c1,c2,c1}
    \end{fmfgraph}
    \end{fmffile}
}
\!-\frac12\;
\unitlength=0.12em
\parbox{37\unitlength}{
    \begin{fmffile}{loop2c}
    \begin{fmfgraph}(30,25)\fmfkeep{loopb}
        \fmfpen{thin}\fmfleftn{l}{2}\fmfrightn{r}{2}
        \fmf{fermion}{l2,r2}\fmf{phantom}{l2,c1,c2,r2}
        \fmf{plain,left,width=thin,tension=.1}{c1,c2,c1}
        \fmf{photon}{r2,r1}\fmf{fermion}{r1,l1}
        \fmf{phantom}{r1,c3,c4,l1}
        \fmf{plain,left,width=thin,tension=.1}{c3,c4,c3}
        \fmf{photon}{l1,l2}
    \end{fmfgraph}
    \end{fmffile}
}
\!-\frac13\;
\unitlength=0.12em
\parbox{37\unitlength}{
    \begin{fmffile}{loop3c}
    \begin{fmfgraph}(30,25)\fmfkeep{loopc}
        \fmfpen{thin}\fmfleftn{l}{3}\fmfrightn{r}{3}
        \fmftopn{t}{4}\fmfbottomn{b}{4}
        \fmf{fermion}{l2,t2}\fmf{photon}{t2,t3}
        \fmf{fermion}{t3,r2}\fmf{photon}{r2,b3}
        \fmf{fermion}{b3,b2}\fmf{photon}{b2,l2}
        \fmf{phantom}{l2,c1,ca,c2,t2}
        \fmf{phantom}{t3,c3,cb,c4,r2}
        \fmf{phantom}{b3,c5,cc,cd,ce,c6,b2}
        \fmffreeze
        \fmf{plain,left,width=thin}{c1,c2,c1}
        \fmf{plain,left,width=thin}{c3,c4,c3}
        \fmf{plain,left,width=thin}{c5,c6,c5}
    \end{fmfgraph}
    \end{fmffile}
}
\!-\dots
\label{eq17}\\
&&\qquad\qquad-\;\;
\unitlength=0.16em
\parbox{44\unitlength}{
    \begin{fmffile}{vert2c}
    \begin{fmfgraph*}(40,25)
        \fmfpen{thin}\fmfleftn{l}{3}\fmfrightn{r}{3}
        \fmfpolyn{label=$\bigcirc$}{p}{4}
        \fmf{photon,right=.7}{p1,r2,p2}\fmf{photon,right=.7}{p3,l2,p4}
    \end{fmfgraph*}
    \end{fmffile}
}
\!-\;
\unitlength=0.12em
\parbox{34\unitlength}{
    \begin{fmffile}{vert3c}
    \begin{fmfgraph*}(30,25)
        \fmfpen{thin}\fmftopn{t}{3}\fmfbottomn{b}{2}
        \fmfpolyn{label=$\bigcirc$}{p}{6}
        \fmf{photon,right=.9}{p1,b2,p2}\fmf{photon,right=.9}{p3,t2,p4}
        \fmf{photon,right=.9}{p5,b1,p6}
    \end{fmfgraph*}
    \end{fmffile}
}
\!-\dots-\;
\unitlength=0.16em
\parbox{44\unitlength}{
    \begin{fmffile}{vert2lc}
    \begin{fmfgraph*}(40,15)
        \fmfpen{thin}\fmfleftn{l}{2}\fmfrightn{r}{3}
        \fmfpolyn{label=$\bigcirc$}{p}{4}
        \fmf{photon,right=.7}{p1,r2,p2}\fmf{photon,right=.3}{p3,l2}
        \fmf{fermion}{l1,l2}\fmf{photon,right=.3}{l1,p4}
        \fmf{phantom}{l1,c1,ca,c2,l2}
        \fmf{plain,width=thin,left,tension=.1}{c1,c2,c1}
    \end{fmfgraph*}
    \end{fmffile}
}
\!-\dots
\left.
\Biggr\}\right\rangle_0,
\nonumber
\end{eqnarray}
where arrows denote the zero-order Green's functions (\ref{eq16}), wavy lines
denote hopping integrals and $\square$, \ldots\ stay for some complicated
``$n$ vertices'', which for such type perturbation expansion are an irreducible
many-particle single-site Green's functions calculated with single-site
Hamiltonian (\ref{Hdiag}). Each vertex (Green's function)
is multiplied by a diagonal Hubbard operator denoted by a circle and one
gets an expression with averages of the products of diagonal Hubbard
operators.

For the Falicov-Kimball model expression (\ref{eq17}) reduces and
contains only single loop contributions
\begin{equation}\label{SigmaFK}
\left\langle\hat\sigma(\beta)\right\rangle_0=\left\langle
\exp\Biggl\{-\;
\unitlength=0.1em
\parbox{45\unitlength}{\fmfreuse{loopa}}
\!-\frac12\;
\unitlength=0.12em
\parbox{35\unitlength}{\fmfreuse{loopb}}
\!-\frac13\;
\unitlength=0.12em
\parbox{34\unitlength}{\fmfreuse{loopc}}
\!-\dots
\Biggr\}\right\rangle_0,
\label{eq18}
\end{equation}
where $\;
\unitlength=0.1em
\parbox{47\unitlength}{\begin{fmffile}{arrow}
    \begin{fmfgraph}(40,10)\fmfkeep{arrow0}
        \fmfpen{thin}\fmfleft{el}\fmfright{er}
        \fmf{fermion}{el,er}\fmf{phantom}{el,c1,c2,er}
        \fmf{plain,left,width=thin,tension=.2}{c1,c2,c1}
    \end{fmfgraph}
    \end{fmffile}}\!
=\frac{\hat P_i^\pm}{i\omega_n+\mu^*{\mp}\frac U2}$;
$\hat{P}_{i}^{+} = \hat{n}_{i\downarrow}$; $\hat{P}^{-} =
1-\hat{n}_{i\downarrow}$, $\mu^*=\mu-\frac U2$
and by introducing pseudospin variables
$S_{i}^{z}=\frac{1}{2} (\hat{P}_{i}^{+} - \hat{P}_{i}^{-})$ one
can transform the Falicov-Kimball model into an Ising-type model with the effective
multisite retarded pseudospin interactions. Expression
(\ref{SigmaFK}) can be obtained from the statistical
operator (\ref{eq1}) by performing partial averaging over
fermionic variables, which gives an effective statistical operator
for pseudospins (ions).

So, after applying Wick's theorem our problem splits into two problems:
(i) calculation of the irreducible many-particle Green's functions (vertices)
in order to construct expression (\ref{eq17}) and
(ii) calculation of the averages of the products of diagonal Hubbard
operators and summing up the resulting series.

\section{Irreducible many-particle Green's functions}

For the Hubbard model by applying the Wick's theorem for $X$ operators one gets for
two-vertex
\begin{equation}
\label{eq19}
\unitlength=0.1em
\parbox{47\unitlength}{\fmfreuse{arrow0}}
= g_{\sigma 0} (\omega_{n}) (\hat{X}_{i}^{\sigma\sigma} + \hat{X}_{i}^{00})
+ g_{2\bar{\sigma}} (\omega_{n}) (\hat{X}_{i}^{22} +
\hat{X}_{i}^{\bar{\sigma}\bar{\sigma}}),
\end{equation}
for four-vertex
\begin{eqnarray}
\unitlength=0.1em
\strut_{\omega_{n+m}\sigma}^{\omega_n\sigma}\parbox{20\unitlength}
{\begin{fmffile}{pure4ver}
\begin{fmfgraph*}(15,15)\fmfkeep{pure4}\fmfstraight\fmfpen{thin}
\fmfleftn{l}{2}\fmfrightn{r}{2}
\fmf{plain}{p1,r1}\fmf{plain}{p2,r2}\fmf{plain}{p3,l2}\fmf{plain}{p4,l1}
\fmfpolyn{label=$\bigcirc$,tension=.2}{p}{4}
\end{fmfgraph*}
\end{fmffile}}\strut_{\omega_{n'+m}\bar\sigma}^{\omega_{n'}\bar\sigma}
&=&
\hat\Lambda_{i\sigma\bar\sigma}^{(4)}(\omega_n,\omega_{n+m},\omega_{n'+m},\omega_{n'})
\nonumber\\
&=& \hat{X}_{i}^{00} g_{\sigma 0}(\omega_{n}) g_{\sigma 0} (\omega_{n+m})
\left(U+ U^{2}g_{20} (\omega_{n+n'+m})\right) g_{\bar{\sigma}0} (\omega_{n'})
g_{\bar{\sigma}0} (\omega_{n'+m})
\nonumber\\
&& +\hat{X}_{i}^{22} g_{2\bar{\sigma}} (\omega_{n}) g_{2\bar{\sigma}}
(\omega_{n+m}) \left(U-U^{2} g_{20}(\omega_{n+n'+m})\right) g_{2\sigma} (\omega_{n'})
g_{2\sigma} (\omega_{n'+m})
\nonumber\\
&& +\hat{X}_{i}^{\sigma\sigma} g_{\sigma 0} (\omega_{n}) g_{\sigma 0}
(\omega_{n+m}) \left(U+U^{2} g_{\sigma\bar{\sigma}} (\omega_{n-n'})\right)
g_{2\sigma}(\omega_{n'}) g_{2\sigma} (\omega_{n'+m})
\label{eq20}\\
&& +\hat{X}_{i}^{\bar{\sigma}\bar{\sigma}} g_{2\bar{\sigma}} (\omega_{n})
g_{2\bar{\sigma}} (\omega_{n+m}) \left(U-U^{2} g_{\sigma\bar{\sigma}}
(\omega_{n-n'})\right) g_{\bar{\sigma}0} (\omega_{n'}) g_{\bar{\sigma}0}
(\omega_{n'+m}),
\nonumber
\end{eqnarray}
\[
\hat\Lambda_{i\sigma\sigma}^{(4)}(\omega_n,\omega_{n+m},\omega_{n'+m},\omega_{n'})
\equiv 0
\]
\btwocol
\narrowtext\noindent
and so on. Expressions (\ref{eq19}) and (\ref{eq20}) and for the vertices of
higher order possess one significant feature. They decompose into four terms
with different diagonal Hubbard operators $X^{pp}$, which project our single-site
problem on certain ``vacuum'' states (subspaces), and zero-order Green's
functions, which describe all possible excitations and scattering processes
around these ``vacuum'' states: i.e., creation and annihilation of
single electrons and of the doublon (pair of electrons with opposite spins)
for subspaces $p=0$ and $p=2$ and creation and
annihilation of single electrons with appropriate spin orientation and of
the magnon (spin flip) for subspaces $p=\uparrow$ and $p=\downarrow$.

In compact form expressions (\ref{eq19}) and (\ref{eq20}) can be written as
\begin{equation}
\unitlength=0.1em
\parbox{47\unitlength}{\fmfreuse{arrow0}}
=\sum_{p} \hat X_{i}^{pp} g_{\sigma(p)}(\omega_{n})
\label{eq21}
\end{equation}
and
\begin{eqnarray}
\unitlength=0.1em
\parbox{20\unitlength}{\fmfreuse{pure4}}&=&
\sum_{p} \hat{X}_{i}^{pp}
g_{\sigma(p)} (\omega_{n}) g_{\sigma(p)} (\omega_{n+m})
\label{eq22}\\
&&\qquad
\times\widetilde{U}_{\sigma\bar{\sigma}(p)} (\omega_{n},\omega_{l}|\omega_{m})
g_{\bar{\sigma}(p)} (\omega_{l}) g_{\bar{\sigma}(p)}
(\omega_{l+m}),
\nonumber
\end{eqnarray}
where
\begin{equation}
g_{\sigma(p)}(\omega_{n}) = \left\{ \begin{array}{ll}
g_{\sigma 0} (\omega_{n}) & \text{ for $p=0,\sigma$} \\
g_{2\bar\sigma}(\omega_{n}) & \text{ for $p=\bar\sigma,2$}
\end{array} \right..
\label{eq23}
\end{equation}
Here
\begin{eqnarray}
\widetilde{U}_{\sigma\bar{\sigma}(p)} (\omega_{n},\omega_{l}|\omega_{m})
&=&
\left\{ \begin{array} {ll}
\!\!U\pm U^2 g_{20}(\omega_{n+l+m}) & \text{for $p=0,2$} \\
\!\!U\pm U^2 g_{\sigma\bar{\sigma}} (\omega_{n-l}) & \text{for
$p=\sigma,\bar{\sigma}$}
\end{array}
\right.
\nonumber\\
\label{eq24}
\widetilde{U}_{\sigma\bar{\sigma}(p)} (\omega_{n},\omega_{l}|\omega_{m})
&=&
\widetilde{U}_{\bar{\sigma}\sigma(p)} (\omega_{l},\omega_{n}|\omega_{m})
\end{eqnarray}
is a renormalized Coulombic interaction in the subspaces. In diagrammatic
notations expressions (\ref{eq20}) or (\ref{eq22}) can be
represented as
\begin{equation}
\bigskip
\unitlength=0.1em\parbox{20\unitlength}
{\begin{fmffile}{pure4vr1}
\begin{fmfgraph*}(15,15)\fmfkeep{pure4}\fmfstraight\fmfpen{thin}
\fmfleftn{l}{2}\fmfrightn{r}{2}
\fmf{plain}{p1,r1}\fmf{plain}{p2,r2}\fmf{plain}{p3,l2}\fmf{plain}{p4,l1}
\fmfpolyn{tension=.2}{p}{4}
\fmflabel{1}{l2}\fmflabel{2}{l1}\fmflabel{3}{r1}\fmflabel{4}{r2}
\end{fmfgraph*}
\end{fmffile}}
\;\;=\;\;
\unitlength=0.1em
\parbox{25\unitlength}{\begin{fmffile}{4vert}
\begin{fmfgraph*}(20,20)\fmfstraight
    \fmfpen{thin}\fmfleftn{l}{2}\fmfrightn{r}{2}
    \fmf{fermion}{l2,c,l1}\fmf{fermion}{r1,c,r2}\fmfdot{c}
    \fmflabel{1}{l2}\fmflabel{2}{l1}\fmflabel{3}{r1}\fmflabel{4}{r2}
\end{fmfgraph*}
\end{fmffile}}
\;\;\pm
\left\{\;\;\begin{array}{ll}
\unitlength=0.1em
\parbox{55\unitlength}{
\begin{fmffile}{4vertd}
\begin{fmfgraph*}(50,20)\fmfstraight
    \fmfpen{thin}\fmfleftn{l}{2}\fmfrightn{r}{2}
    \fmf{fermion}{l2,c1}\fmf{fermion}{l1,c1}\fmf{fermion}{c2,r1}
    \fmf{fermion}{c2,r2}\fmf{scalar}{c1,c2}\fmfdotn{c}{2}
    \fmflabel{1}{l2}\fmflabel{3}{l1}\fmflabel{4}{r1}\fmflabel{2}{r2}
\end{fmfgraph*}
\end{fmffile}} & \text{ for $p=0,2$}\\ \\ \\
\parbox{55\unitlength}{
\unitlength=0.1em
\begin{fmffile}{4vertm}
\begin{fmfgraph*}(50,20)\fmfstraight
    \fmfpen{thin}\fmfleftn{l}{2}\fmfrightn{r}{2}
    \fmf{fermion}{l2,c1,l1}\fmf{fermion}{r1,c2,r2}
    \fmf{scalar}{c1,c2}\fmfdotn{c}{2}
    \fmflabel{1}{l2}\fmflabel{4}{l1}\fmflabel{3}{r1}\fmflabel{2}{r2}
\end{fmfgraph*}
\end{fmffile}} & \text{ for $p=\sigma,\bar\sigma$}
\end{array}
\right.,
\bigskip
\bigskip
\label{eq25}
\end{equation}
where dots denote Coulombic correlation energy $U=\lambda_{2}+\lambda_{0} -
\lambda_{\uparrow} - \lambda_{\downarrow}$ and dashed arrows denote bosonic
zero-order Green's functions: doublon $g_{20}(\omega_{m})$ or magnon
$g_{\sigma\bar{\sigma}}(\omega_{m})$.

For six-vertex one can get
\etwocol
\widetext
\[
\hat\Lambda_{i\sigma\sigma\sigma}^{(6)}(\omega_n,\omega_{n_1},\omega_{n_2},
\omega_{n_3},\omega_{n_4},\omega_{n_5})
\equiv 0,
\]
\begin{eqnarray}
\hat\Lambda_{i\sigma\bar\sigma\bar\sigma}^{(6)}(\omega_n,\omega_{n_1},\omega_{n_2},
\omega_{n_3},\omega_{n_4},\omega_{n_5})=\delta(\omega_n-\omega_{n_1}+\omega_{n_2}-
\omega_{n_3}+\omega_{n_4}-\omega_{n_5})
\nonumber\\
\times\sum_p \hat X_i^{pp}
g_{\sigma(p)}(\omega_n)g_{\sigma(p)}(\omega_{n_1})
g_{\bar\sigma(p)}(\omega_{n_2})g_{\bar\sigma(p)}(\omega_{n_3})
g_{\bar\sigma(p)}(\omega_{n_4})g_{\bar\sigma(p)}(\omega_{n_5})
\nonumber\\
\times\Bigl\{
\widetilde{U}_{\sigma\bar{\sigma}(p)}(\omega_{n},\omega_{n_3}|\omega_{n_2-n_3})
g_{\bar\sigma(p)}(\omega_{n+n_2-n_3})
\widetilde{U}_{\sigma\bar{\sigma}(p)}(\omega_{n_1},\omega_{n_4}|\omega_{n_5-n_4})
\label{eq25p}\\
-\widetilde{U}_{\sigma\bar{\sigma}(p)}(\omega_{n},\omega_{n_5}|\omega_{n_2-n_5})
g_{\bar\sigma(p)}(\omega_{n+n_2-n_5})
\widetilde{U}_{\sigma\bar{\sigma}(p)}(\omega_{n_1},\omega_{n_4}|\omega_{n_3-n_4})
\nonumber\\
-\widetilde{U}_{\sigma\bar{\sigma}(p)}(\omega_{n},\omega_{n_3}|\omega_{n_4-n_3})
g_{\bar\sigma(p)}(\omega_{n+n_4-n_3})
\widetilde{U}_{\sigma\bar{\sigma}(p)}(\omega_{n_1},\omega_{n_2}|\omega_{n_5-n_2})
\nonumber\\
+\widetilde{U}_{\sigma\bar{\sigma}(p)}(\omega_{n},\omega_{n_5}|\omega_{n_4-n_5})
g_{\bar\sigma(p)}(\omega_{n+n_4-n_5})
\widetilde{U}_{\sigma\bar{\sigma}(p)}(\omega_{n_1},\omega_{n_2}|\omega_{n_3-n_2})
\nonumber\\
+\Upsilon_{\sigma\bar\sigma\bar\sigma(p)}(\omega_n,\omega_{n_1},\omega_{n_2},
\omega_{n_3},\omega_{n_4},\omega_{n_5})
\Bigr\},
\nonumber
\end{eqnarray}
where
\begin{eqnarray}
\lefteqn{\Upsilon_{\sigma\bar\sigma\bar\sigma(p)}(\omega_n,\omega_{n_1},\omega_{n_2},
\omega_{n_3},\omega_{n_4},\omega_{n_5})}
\label{eq25pp}\\
&& =\left\{\begin{array}{ll}
\pm U^3\left(g_{20}(\omega_{n+n_2})-g_{20}(\omega_{n+n_4})\right)
\left(g_{20}(\omega_{n_1+n_3})-g_{20}(\omega_{n_1+n_5})\right) &
\text{ for $p=0,2$} \\
\pm U^3\left(g_{\sigma\bar\sigma}(\omega_{n-n_3})-g_{\sigma\bar\sigma}(\omega_{n-n_5})\right)
\left(g_{\sigma\bar\sigma}(\omega_{n_1-n_2})-g_{\sigma\bar\sigma}(\omega_{n_1-n_4})\right) &
\text{ for $p=\sigma,\bar\sigma$}
\end{array}
\right..
\nonumber
\end{eqnarray}
\btwocol
\narrowtext\noindent
In expression (\ref{eq25p}) the contributions of the first four terms in
braces can be presented by the following diagrams:
\begin{equation}
\unitlength=0.1em
\begin{fmffile}{6vert0}
\begin{fmfgraph}(40,20)\fmfstraight
    \fmfpen{thin}\fmfleft{l}\fmfright{r}\fmftopn{t}{5}\fmfbottomn{b}{5}
    \fmf{plain}{l,c1,c2,r}\fmf{plain}{b2,c1,t2}\fmf{plain}{b4,c2,t4}
    \fmfdotn{c}{2}
\end{fmfgraph}
\quad
\begin{fmfgraph}(40,20)\fmfstraight
    \fmfpen{thin}\fmfleft{l}\fmfrightn{r}{4}\fmftopn{t}{5}\fmfbottomn{b}{5}
    \fmf{plain}{l,c1,c2,t3}\fmf{plain}{b2,c1,t2}\fmf{plain}{b4,c3,r2}
    \fmf{dashes}{c2,c3}\fmfdotn{c}{3}
\end{fmfgraph}
\quad
\begin{fmfgraph}(40,20)\fmfstraight
    \fmfpen{thin}\fmfleftn{l}{4}\fmfrightn{r}{4}\fmftopn{t}{6}\fmfbottomn{b}{6}
    \fmf{plain}{l3,c1,t2}\fmf{plain}{b3,c2,c3,b4}\fmf{plain}{t5,c4,r3}
    \fmf{dashes}{c1,c2}\fmf{dashes}{c3,c4}\fmfdotn{c}{4}
\end{fmfgraph}
\end{fmffile}
\label{eq26}
\end{equation}
with the internal vertices of the same type as in Eq.~(\ref{eq25}), whereas the
contribution of the last term can be presented diagrammatically as
\begin{equation}
\unitlength=0.1em
\begin{fmffile}{6vert1}
\begin{fmfgraph}(40,20)\fmfstraight
    \fmfpen{thin}\fmfleftn{l}{2}\fmfrightn{r}{2}\fmftopn{t}{4}
    \fmf{plain}{l1,c1,l2}\fmf{plain}{r1,c3,r2}
    \fmf{dashes}{c1,c2,c3}\fmfdotn{c}{3}
    \fmffreeze\fmf{plain}{t2,c2,t3}
\end{fmfgraph}
\end{fmffile}
\label{eq27}
\end{equation}
So, we can introduce primitive vertices
\begin{equation}
\unitlength=0.1em
\begin{fmffile}{element}
\begin{fmfgraph}(20,20)\fmfstraight
    \fmfpen{thin}\fmfleftn{l}{2}\fmfrightn{r}{2}
    \fmf{plain}{l1,c,l2}\fmf{plain}{r1,c,r2}\fmfdot{c}
\end{fmfgraph}
\quad
\begin{fmfgraph}(20,20)\fmfstraight
    \fmfpen{thin}\fmfleftn{l}{2}\fmfright{r}
    \fmf{plain}{l1,c,l2}\fmf{dashes}{c,r}\fmfdot{c}
\end{fmfgraph}
\quad
\begin{fmfgraph}(20,20)\fmfstraight
    \fmfpen{thin}\fmfleftn{l}{2}\fmfrightn{r}{2}
    \fmf{plain}{l1,c,l2}\fmf{dashes}{r1,c,r2}\fmfdot{c}
\end{fmfgraph}
\end{fmffile}
\label{eq28}
\end{equation}
by which one can construct all $n$ vertices in expansion (\ref{eq17})
according to the following rules:
\begin{enumerate}
\item $n$ vertices are constructed by the diagonal Hubbard operator $X^{pp}$
and zero-order fermionic and bosonic
lines connected by primitive vertices (\ref{eq28}) specific for
each subspace $p$.
\item External lines of $n$ vertices must be of the fermionic
type.
\item Diagrams with the loops formed by zero-order fermionic and bosonic
Green's functions are not allowed because they are already included into the
formalism, e.g.,
\(
\unitlength=0.05em
\begin{fmffile}{restrict}
\begin{fmfgraph}(40,20)\fmfstraight
    \fmfpen{thin}\fmfleftn{l}{2}\fmfrightn{r}{2}
    \fmf{plain}{l1,c1,l2}\fmf{plain}{r1,c2,r2}
    \fmf{plain,left=.5,tension=.3}{c1,c2,c1}
    \fmfdotn{c}{2}
\end{fmfgraph}
\quad\text{ gives }\quad
\begin{fmfgraph}(40,20)\fmfstraight
    \fmfpen{thin}\fmfleftn{l}{2}\fmfrightn{r}{2}
    \fmf{plain}{l1,c1,l2}\fmf{plain}{r1,c2,r2}
    \fmf{dashes,tension=.5}{c1,c2}
    \fmfdotn{c}{2}
\end{fmfgraph}
\end{fmffile}
\)
\end{enumerate}
For $n$ vertices of higher
order a new primitive vertices can appear but we do not check this due to
the rapid increase of the algebraic calculations with the increase of $n$.
Diagrams (\ref{eq25}), (\ref{eq26}), and (\ref{eq27}) topologically
are truncated Bethe-lattices constructed by the primitive vertices
(\ref{eq28}) and can be treated as some generalization of the
Hubbard stars\cite{Dongen,Gros} in the thermodynamical perturbation theory.

It should be noted that each $n$ vertex contains Coulombic
interaction $U$ as in primitive vertices (\ref{eq28}) (denoted by dots) as in
the denominators of the zero-order Green's functions (\ref{eq16}).
In the $U\to\infty$ limit, each term in the expressions for $n$ vertices
can diverge but total vertex possesses finite $U\to\infty$ limit
when diagrammatic series of
Ref.~\onlinecite{IzyumovLetfulov} are reproduced.

The second problem of calculation of the averages of diagonal $X$ operators
is more complicated. One of the ways to solve it is to use semi-invariant
(cumulant) expansions as was done in Refs.~\onlinecite{IzyumovLetfulov}
and \onlinecite{IzyumovPRB}
for the $U=\infty$ limit. Another way
is to consider the $d=\infty$ limit where new simplifications appear.

\section{Dynamical mean-field theory}

Within the frames of the considered perturbation theory in terms of electron
hopping a single-electron Green's function (\ref{eq9}) can be presented in a
form
\begin{equation}
G_{\sigma}(\omega_{n},\boldsymbol{k}) = \frac{1}{\Xi_{\sigma}^{-1}
(\omega_{n},\boldsymbol{k}) - t_{\boldsymbol{k}}},
\label{eq29}
\end{equation}
where we introduce an irreducible part
$\Xi_{\sigma}(\omega_{n},\boldsymbol{k})$ of Green's
function which, in general, is not local. In the case of infinite dimensions
$d\rightarrow\infty$ one should scale the hopping integral according to
\begin{equation}
t_{ij}\rightarrow \frac{t_{ij}}{\sqrt{d}}
\end{equation}
in order to obtain finite density-of-states and it was shown by Metzner in
his pioneer work\cite{Metzner} that in this limit the irreducible part
become local
\begin{equation}
\Xi_{ij\sigma} (\tau-\tau')= \delta_{ij} \Xi_{\sigma} (\tau-\tau')
\quad\text{ or }\quad
\Xi_{\sigma}(\omega_{n},\boldsymbol{k}) = \Xi_{\sigma} (\omega_{n})
\label{eq30}
\end{equation}
and such a site-diagonal function, as it was shown by Brandt and
Mielsch,\cite{BrandtMielsch}
can be calculated by mapping the infinite-dimensional lattice problem
(\ref{eq1}) with $t_{ij}^{\sigma}(\tau-\tau') =
\frac{1}{\sqrt{d}}t_{ij}\delta(\tau-\tau')$ on
the atomic model with auxiliary the Kadanoff-Baym field
\begin{equation}
t_{ij}^{\sigma}(\tau-\tau') = \delta_{ij} J_{\sigma} (\tau-\tau'),
\label{eq31}
\end{equation}
which has to be self-consistently determined from the condition that the
same function $\Xi_{\sigma}(\omega_{n})$ defines Green's functions for the
lattice and atomic limit. The self-consistent set of equations for
$\Xi_{\sigma}(\omega_{n})$ and $J_{\sigma}(\omega_{n})$
(e.g., see Ref.~\onlinecite{DMFTreview} and references therein) is the following:
\begin{eqnarray}
\frac{1}{N} \sum_{\boldsymbol{k}} \frac{1}{\Xi_{\sigma}^{-1}(\omega_{n})
-t_{\boldsymbol{k}}} &=& \frac{1}{\Xi_{\sigma}^{-1}(\omega_{n}) -
J_{\sigma}(\omega_{n})}
\nonumber\\
&=& G_{\sigma}^{(a)} (\omega_{n}, \{ J_{\sigma}
(\omega_{n})\}),
\label{eq32}
\end{eqnarray}
where $G_{\sigma}^{(a)}(\omega_{n},\{J_{\sigma}(\omega_{n})\})$ is a Green's
function for atomic limit (\ref{eq31}).

Grand canonical potential for lattice is connected
with the one for atomic limit by the expression\cite{BrandtMielsch}
\begin{equation}
\frac{\Omega}{N} = \Omega_{a} - \frac1{\beta} \sum_{n\sigma} \left\{ \ln
G_{\sigma}^{(a)} (\omega_{n}) - \frac{1}{N} \sum_{\boldsymbol{k}} \ln G_{\sigma}
(\omega_{n},\boldsymbol{k})\right\}.
\label{eq33}
\end{equation}

On the other hand, we can write for the grand canonical potential for atomic
limit $\Omega_{a}$ the same expansion as in Eq.~(\ref{eq17}) but now we have
averages of the products of diagonal $X$ operators at the same site.
According to Eq.~(\ref{eq12}) we can multiply them and reduce their product to a
single $X$ operator that can be taken outside of the brackets and exponent
in (\ref{eq17}) and its average is equal to $\langle X^{pp}\rangle_{0} =
\frac{e^{-\beta\lambda_{p}}}{\sum_q e^{-\beta\lambda_{q}}}$.
Finally, for the grand canonical potential for atomic limit we get
\begin{equation}
\Omega_{a} = -\frac{1}{\beta} \ln \sum_{p} e^{-\beta\Omega_{(p)}},
\label{eq34}
\end{equation}
where
\etwocol
\widetext
\begin{eqnarray}
\Omega_{(p)}&=&
\lambda_p+\frac1{\beta}\Biggl\{\;
\unitlength=0.1em
\parbox{47\unitlength}{
    \begin{fmffile}{loop1}
    \begin{fmfgraph}(40,25)
        \fmfpen{thin}\fmfleft{el}\fmfright{er}
        \fmf{photon,right=.7}{er,el}
        \fmf{fermion}{er,el}
    \end{fmfgraph}
    \end{fmffile}
}
\!+\frac12\;
\unitlength=0.12em
\parbox{36\unitlength}{
    \begin{fmffile}{loop2}
    \begin{fmfgraph}(30,25)
        \fmfpen{thin}\fmfleftn{l}{2}\fmfrightn{r}{2}
        \fmf{fermion}{l2,r2}\fmf{photon}{r2,r1}
        \fmf{fermion}{r1,l1}\fmf{photon}{l1,l2}
    \end{fmfgraph}
    \end{fmffile}
}
\!+\frac13\;
\unitlength=0.12em
\parbox{36\unitlength}{
    \begin{fmffile}{loop3}
    \begin{fmfgraph}(30,25)
        \fmfpen{thin}\fmfleftn{l}{3}\fmfrightn{r}{3}
        \fmftopn{t}{4}\fmfbottomn{b}{4}
        \fmf{fermion}{l2,t2}\fmf{photon}{t2,t3}
        \fmf{fermion}{t3,r2}\fmf{photon}{r2,b3}
        \fmf{fermion}{b3,b2}\fmf{photon}{b2,l2}
    \end{fmfgraph}
    \end{fmffile}
}
\!+\dots
\label{eq35}\\
&&\qquad\qquad+\;\;
\unitlength=0.16em
\parbox{44\unitlength}{
    \begin{fmffile}{vert2}
    \begin{fmfgraph}(40,25)
        \fmfpen{thin}\fmfleftn{l}{3}\fmfrightn{r}{3}\fmfpolyn{empty}{p}{4}
        \fmf{photon,right=.7}{p1,r2,p2}\fmf{photon,right=.7}{p3,l2,p4}
    \end{fmfgraph}
    \end{fmffile}
}
\!+\;
\unitlength=0.12em
\parbox{35\unitlength}{
    \begin{fmffile}{vert3}
    \begin{fmfgraph}(30,25)
        \fmfpen{thin}\fmftopn{t}{3}\fmfbottomn{b}{2}
        \fmfpolyn{empty}{p}{6}\fmf{photon,right=.9}{p1,b2,p2}
        \fmf{photon,right=.9}{p3,t2,p4}\fmf{photon,right=.9}{p5,b1,p6}
    \end{fmfgraph}
    \end{fmffile}
}
\!+\dots+\;
\unitlength=0.16em
\parbox{44\unitlength}{
    \begin{fmffile}{vert2l}
    \begin{fmfgraph}(40,15)
        \fmfpen{thin}\fmfleftn{l}{2}\fmfrightn{r}{3}\fmfpolyn{empty}{p}{4}
        \fmf{photon,right=.7}{p1,r2,p2}\fmf{photon,right=.3}{p3,l2}
        \fmf{fermion}{l1,l2}\fmf{photon,right=.3}{l1,p4}
    \end{fmfgraph}
    \end{fmffile}
}
\!+\dots
\Biggr\}
\nonumber
\end{eqnarray}
\btwocol
\narrowtext\noindent
are the ``grand canonical potentials'' for the subspaces.

Now we can find single-electron Green's function for atomic limit by
\begin{equation}
G_{\sigma}^{(a)} (\tau-\tau') = \frac{\delta\Omega_{a}}{\delta
J_{\sigma}(\tau-\tau')} = \sum_{p} w_{p} G_{\sigma(p)} (\tau-\tau'),
\label{eq36}
\end{equation}
where
\begin{equation}
G_{\sigma(p)} (\tau-\tau') = \frac{\delta\Omega_{(p)}}{\delta
J_{\sigma}(\tau-\tau')}
\label{eq37}
\end{equation}
are single-electron Green functions for the subspaces characterized by
the ``statistical weights''
\begin{equation}
w_{p} = \frac{e^{-\beta\Omega_{(p)}}}{\sum\limits_{q}e^{-\beta\Omega_{(q)}}}
\label{eq38}
\end{equation}
and our single-site atomic problem exactly (naturally) splits into four
subspaces $p=0,2,\downarrow,\uparrow$.

We can introduce irreducible parts of Green's functions in subspaces
$\Xi_{\sigma(p)}(\omega_{n})$ by
\begin{equation}
G_{\sigma(p)} (\omega_{n}) = \frac{1}{\Xi_{\sigma(p)}^{-1}(\omega_{n}) -
J_{\sigma}(\omega_{n})},
\label{eq39}
\end{equation}
where
\begin{equation}
\unitlength=0.12em
\begin{fmffile}{irreduc}
\Xi_{\sigma(p)}(\omega_n)=
\begin{fmfgraph}(20,5)\fmfstraight
    \fmfpen{thin}\fmfleft{l}\fmfright{r}\fmf{fermion}{l,r}
\end{fmfgraph}
+
\begin{fmfgraph}(20,15)\fmfstraight
    \fmfpen{thin}\fmfbottomn{b}{5}\fmftop{t}
    \fmf{plain}{b2,p4}\fmf{plain}{p1,b4}\fmfpolyn{tension=.1}{p}{4}
    \fmffreeze
    \fmf{photon,right=.7}{p2,t,p3}
\end{fmfgraph}
+
\begin{fmfgraph}(20,15)\fmfstraight
    \fmfpen{thin}\fmfbottomn{b}{5}\fmftopn{t}{3}
    \fmf{plain}{b2,p6}\fmf{plain}{p1,b4}\fmfpolyn{tension=.1}{p}{6}
    \fmffreeze
    \fmf{photon,right=.7}{p2,t3,p3}\fmf{photon,right=.7}{p4,t1,p5}
\end{fmfgraph}
+\ldots.
\end{fmffile}
\end{equation}

According to the rules of the introduced diagrammatic technique, $n$ ver\-ti\-ces
are terminated by the fermionic Green's functions [see (\ref{eq25}),
(\ref{eq26}), and
(\ref{eq27})] and this allows us to write a Dyson equation for the irreducible parts
and to introduce a self-energy in subspaces
\begin{equation}
\Xi_{\sigma(p)}^{-1} (\omega_{n}) = g_{\sigma(p)}^{-1} (\omega_{n}) -
\Sigma_{\sigma(p)} (\omega_{n}),
\label{eq40}
\end{equation}
where self-energy $\Sigma_{\sigma(p)}(\omega_{n})$ depends on the hopping
integral $J_{\sigma'}(\omega_{n'})$ only through quantities
\begin{eqnarray}
\lefteqn{\Psi_{\sigma'(p)} (\omega_{n'}) =
G_{\sigma'(p)} (\omega_{n'}) - \Xi _{\sigma'(p)} (\omega_{n'})}
\label{eq41}\\
&&\quad\equiv
\Xi^2_{\sigma'(p)} (\omega_{n'})J_{\sigma'}(\omega_{n'})
\left\{ 1 +
 \Xi_{\sigma'(p)}(\omega_{n'})J_{\sigma'}(\omega_{n'})
+ \cdots
\right\}.
\nonumber
\end{eqnarray}
It should be noted, that the total self-energy of the atomic
problem is connected with the total irreducible part by the
expression
\begin{equation}
\Sigma_{\sigma}(\omega_n)=i\omega_n+\mu-\Xi_{\sigma}^{-1}(\omega_n)
\end{equation}
and it has no direct connection with the self-energies in the
subspaces.

The fermionic zero-order Green's function (\ref{eq23}) can be also
represented in the following form
\begin{equation}
g_{\sigma(p)}=\frac1{i\omega_n+\mu_{\sigma}-Un^{(0)}_{\bar\sigma(p)}},
\end{equation}
where
\begin{equation}
n^{(0)}_{\sigma(p)}=-\frac{d\lambda_p}{d\mu_{\sigma}}=\left\{\begin{array}{cc}
  0 & \text{for } p=0,\bar\sigma \\
  1 & \text{for } p=2,\sigma
\end{array}\right.
\end{equation}
is an occupation of the state $|p\rangle$ by the electron with
spin $\sigma$, and Green's function (\ref{eq39}) can be written as
\begin{equation}\label{GFp}
G_{\sigma(p)}(\omega_n)=\frac1{i\omega_n+\mu_{\sigma}-Un^{(0)}_{\bar\sigma(p)}
-\Sigma_{\sigma(p)}(\omega_n)-J_{\sigma}(\omega_n)}.
\end{equation}

Now, one can reconstruct expressions for the grand canonical potentials
$\Omega_{(p)}$ in subspaces from the known structure of Green's functions.
To do this, we scale hopping integral
\begin{equation}
J_{\sigma}(\omega_{n}) \rightarrow \alpha J_{\sigma}(\omega_{n}),
\;
\alpha\in [0,1],
\end{equation}
which allows to define the grand canonical potential as
\begin{equation}
\Omega_{(p)} = \lambda_{p} + \int\limits_{0}^{1} d\alpha \frac{1}{\beta}
\sum_{n\sigma} J_{\sigma} (\omega_{n}) G_{\sigma(p)} (\omega_{n}, \alpha)
\label{eq42}
\end{equation}
and after some transformations one can get
\begin{eqnarray}
\Omega_{(p)} &=& \lambda_{p} - \frac{1}{\beta} \sum_{n\sigma}
\ln\frac{\Xi_{\sigma(p)}^{-1}(\omega_{n}) - J_{\sigma}(\omega_{n})}
{\Xi_{\sigma(p)}^{-1}(\omega_{n})}
\nonumber\\
&&- \frac{1}{\beta} \sum_{n\sigma} \Sigma_{\sigma(p)}(\omega_{n})
\Psi_{\sigma(p)}(\omega_{n})
+\Phi_{(p)},
\label{eq43}
\end{eqnarray}
where
\begin{equation}\label{Phi}
\Phi_{(p)}=\frac{1}{\beta}
\sum_{n\sigma}\int\limits_0^1\!\!d\alpha\, \Sigma_{\sigma(p)}(\omega_{n},\alpha)
\frac{d\Psi_{\sigma(p)}(\omega_{n},\alpha)}{d\alpha}
\end{equation}
is some functional, such that its functional derivative with
respect to $\Psi$ produces self-energy:
\begin{equation}
\frac{\delta\Phi_{(p)}}{\delta\Psi_{\sigma(p)}(\omega_{n})}
=\Sigma_{\sigma(p)}(\omega_{n}).
\end{equation}
So, if one can find or construct
self-energy $\Sigma_{\sigma(p)}(\omega_{n})$ he can find Green's functions
and grand canonical potentials for subspaces and, according to Eqs.~(\ref{eq34})
and (\ref{eq36}), solve atomic problems.

Starting from the grand canonical potential (\ref{eq34}) and (\ref{eq43}) one can
get for mean values (\ref{eq10}),
\begin{eqnarray}\label{mv}
n_{\sigma}&=&\sum_p w_p n_{\sigma(p)},
\\ \nonumber
n_{\sigma(p)}&=&n^{(0)}_{\sigma(p)}+
\frac1{\beta}\sum_n\left[G_{\sigma(p)}(\omega_{n})-\Xi_{\sigma(p)}(\omega_{n})\right]
-\frac{\partial\Phi_{(p)}}{\partial\mu_{\sigma}},
\end{eqnarray}
where in the last term the partial derivative is taken over the
$\mu_{\sigma}$ not in the chains (\ref{eq41}).
The second term in the right-hand side of Eq.~(\ref{mv}) can be
represented diagrammatically as
\begin{equation}\label{mvPsi}
\unitlength.2em
\begin{fmffile}{mvPsi}
\begin{fmfgraph}(5,10)\fmfstraight
    \fmfpen{thin}\fmftop{t1,t2}\fmfbottom{b}
    \fmf{dbl_plain,left=.4,tension=.5}{b,t1,t2,b}
\end{fmfgraph}
\end{fmffile}
\end{equation}
and the first contributions into the last term are following
\smallskip
\begin{equation}\label{dPhi}
\unitlength.2em
\begin{fmffile}{dPhi}
\begin{fmfgraph}(10,10)\fmfstraight
    \fmfpen{thin}\fmftop{t1,t2}\fmfbottom{b}\fmf{dashes}{t1,b,t2}
    \fmf{dbl_plain,left=.5,tension=.5}{t1,t2}
    \fmf{dbl_plain,right=.5,tension=.5}{t1,t2}
    \fmfdot{t1,t2}
\end{fmfgraph}\quad,\quad
\begin{fmfgraph}(10,10)\fmfstraight
    \fmfpen{thin}\fmftop{t1,t2}\fmfbottom{b}\fmf{plain}{t1,b,t2}
    \fmf{dbl_plain,left=.5,tension=.5}{t1,t2}
    \fmf{dbl_plain,right=.5,tension=.5}{t1,t2}
    \fmf{dbl_plain,tension=.5}{t1,t2}
    \fmfdot{t1,t2}
\end{fmfgraph}\quad, \ldots ,
\end{fmffile}
\end{equation}
where double lines denote quantities $\Psi_{\sigma(p)}(\omega_n)$.
Loop
\unitlength.2em
\begin{fmffile}{bosloop}
\raisebox{.3em}[.8em][.2em]{\begin{fmfgraph}(10,5)\fmfstraight
    \fmfpen{thin}\fmfbottom{l1,l2}
    \fmf{dbl_plain,left=.5,tension=.5}{l1,l2}
    \fmf{dbl_plain,right=.5,tension=.5}{l1,l2}
\end{fmfgraph}}
\end{fmffile}
is connected with the superconducting or magnon susceptibilities
for subspaces $p=0,2$ or $p=\sigma,\bar\sigma$, respectively.

For the single atom
[$J_{\sigma}(\omega_n)=0$] we have $\Phi_{(p)}=0$,
$G_{\sigma(p)}(\omega_n)=\Xi_{\sigma(p)}(\omega_n)=g_{\sigma(p)}(\omega_n)$,
and
\begin{equation}
n_{\sigma}=\sum_p w_p \frac1{\beta}\sum_n g_{\sigma(p)}(\omega_n)
=\sum_p w_p n^{(0)}_{\sigma(p)},
\end{equation}
but in the general case [$J_{\sigma}(\omega_n)\ne0$] we cannot prove
that the sum rule
\begin{equation}
n_{\sigma}=\frac1{\beta}\sum_n G^{(a)}_{\sigma}(\omega_n)
\end{equation}
is fulfilled.

\subsection{Falicov-Kimball model}

For the Falicov-Kimball model $J_{\downarrow}(\omega_n)=0$ and
according to Eqs.~(\ref{eq20}) and (\ref{eq25p}),
\begin{equation}
\Sigma_{\uparrow(p)}(\omega_{n}) \equiv 0; \,\,\,\,
\Xi_{\uparrow(p)}(\omega_{n}) = g_{\uparrow(p)} (\omega_{n})
\label{eq44}
\end{equation}
and
\begin{equation}
\Omega_{(p)} = \lambda_{p} - \frac{1}{\beta} \sum_{n}\ln
\left[1 - J_{\uparrow}(\omega_{n})g_{\uparrow(p)}(\omega_{n})\right],
\label{eq45}
\end{equation}
\begin{equation}
G_{\uparrow}^{(a)}(\omega_n)=\frac{1-n_{\downarrow}}
{i\omega_n-\lambda_{\uparrow0}-J_{\uparrow}(\omega_n)}+
\frac{n_{\downarrow}}
{i\omega_n-\lambda_{2\downarrow}-J_{\uparrow}(\omega_n)},
\label{eq45p}
\end{equation}
\begin{equation}
n_{\uparrow}=\frac1{\beta}\sum_n G_{\uparrow}^{(a)}(\omega_n),
\; n_{\downarrow}=w_2+w_{\downarrow},
\label{eq45pp}
\end{equation}
which immediately gives results of Ref.~\onlinecite{BrandtMielsch}
(see also Ref.~\onlinecite{ShvaikaJPS}).

For the Hubbard model there are no exact expression for self-energy but the set
of Eqs.~(\ref{eq39}), (\ref{eq40}), and (\ref{eq43}) allows one to construct
different self-consistent approximations.

\subsection{Alloy-analogy approximation}

The simplest approximation, which can be done, is to put
\begin{equation}
\Sigma_{\sigma(p)}(\omega_{n})=0
\end{equation}
which gives
\begin{equation}
\Xi_{\sigma(p)}(\omega_{n}) = g_{\sigma(p)} (\omega_{n})
\end{equation}
and
\begin{equation}
\Omega_{(p)} = \lambda_{p} - \frac{1}{\beta} \sum_{n\sigma}\ln
\left[1 - J_{\sigma}(\omega_{n})g_{\sigma(p)}(\omega_{n})\right]
\end{equation}
and for the Green's function
for the atomic problem one can obtain a two-pole expression
\begin{equation}\label{AAA}
G_{\sigma}^{(a)}(\omega_n)=\frac{w_0+w_{\sigma}}
            {i\omega_n-\lambda_{\sigma0}-J_{\sigma}(\omega_n)}+
            \frac{w_2+w_{\bar\sigma}}
            {i\omega_n-\lambda_{2\bar\sigma}-J_{\sigma}(\omega_n)}
\end{equation}
of the alloy-analogy solution for the Hubbard model, which is a
zero-order approximation within the considered approach and is
exact
for the Falicov-Kimball model. For this approximation, mean values (\ref{eq10}) are
equal to
\begin{eqnarray}
n_{\sigma} &=& \frac1{\beta} \sum_n G_{\sigma}^{(a)}(\omega_n) +
w_2+w_{\sigma}-\frac{w_0+w_{\sigma}}{e^{\beta\lambda_{\sigma0}}+\!1}-
\frac{w_2+w_{\bar\sigma}}{e^{\beta\lambda_{2\bar\sigma}}+\!1}
\nonumber\\
&\neq& \frac1{\beta} \sum_n G_{\sigma}^{(a)}(\omega_n)
\end{eqnarray}
and, for some values of the chemical potential, they can get unphysical
values: negative or greater then one.

\subsection{Hartree-Fock approximation}

The next possible approximation is to take into account
the contribution from diagram (\ref{mvPsi}) and to
construct the equation
for the self-energy in the following form:
\begin{equation}
\Sigma_{\sigma(p)}(\omega_n)=\frac1{\beta}\sum_{n'}
U\Psi_{\bar\sigma(p)}(\omega_{n'}),
\end{equation}
which, together with the expression for mean values
\begin{eqnarray}\label{mvHF}
n_{\sigma(p)} &=& n^{(0)}_{\sigma(p)} + \frac{1}{\beta} \sum\limits_{n}
\Psi_{\sigma(p)}(\omega_{n})
\\ \nonumber
&=& n^{(0)}_{\sigma(p)}-\frac12+
\frac12\tanh\frac{\beta}2 \left[Un_{\bar\sigma(p)}-\mu_{\sigma}\right]
\\ \nonumber
&&+\frac1{\beta}\sum_{n'}
G_{\sigma(p)}(\omega_{n'}),
\end{eqnarray}
gives for the Green's function in the subspaces expression in the
Hartree-Fock approximation:
\begin{equation}\label{GFHF}
G_{\sigma(p)}(\omega_n)=\frac1{i\omega_n+\mu_{\sigma}-Un_{\bar\sigma(p)}
-J_{\sigma}(\omega_n)}.
\end{equation}
Now, grand canonical potentials in the subspaces are equal
\begin{eqnarray}
\Omega_{(p)}&=&\lambda_p-\frac1{\beta}\sum_{n\sigma}
\ln\left[1-J_{\sigma}(\omega_n)\Xi_{\sigma(p)}(\omega_n)\right]
\nonumber\\
&&-U \left(n_{\sigma(p)}-n_{\sigma(p)}^{(0)}\right)
\left(n_{\bar\sigma(p)}-n_{\bar\sigma(p)}^{(0)}\right)
\end{eqnarray}
and for the Green's function for the atomic problem (\ref{eq36}) one
can obtain a four-pole structure
\begin{equation}\label{GFHFt}
G^{(a)}_{\sigma}(\omega_n)=\sum_p
\frac{w_p}{i\omega_n+\mu_{\sigma}-Un_{\bar\sigma(p)}
-J_{\sigma}(\omega_n)}.
\end{equation}
Expression (\ref{GFHFt}), in contrast to the alloy-analogy
solution (\ref{AAA}), possesses the correct Hartree-Fock limit for
small Coulombic interaction $U\ll t$:
\begin{equation}
G^{(a)}_{\sigma}(\omega_n)=
\frac1{i\omega_n+\mu_{\sigma}-Un_{\bar\sigma}
-J_{\sigma}(\omega_n)},
\end{equation}
when $w_p\approx\frac14$ and $n_{\sigma(p)}\approx n_{\sigma}=
\frac1{\beta}\sum_n G^{(a)}_{\sigma}(\omega_n)$. On the other
hand, in the same way as an alloy-analogy solution, it describes
the metal-insulator transition with the change of $U$.

\begin{figure}
\noindent\null\hfill\includegraphics[width=60mm]{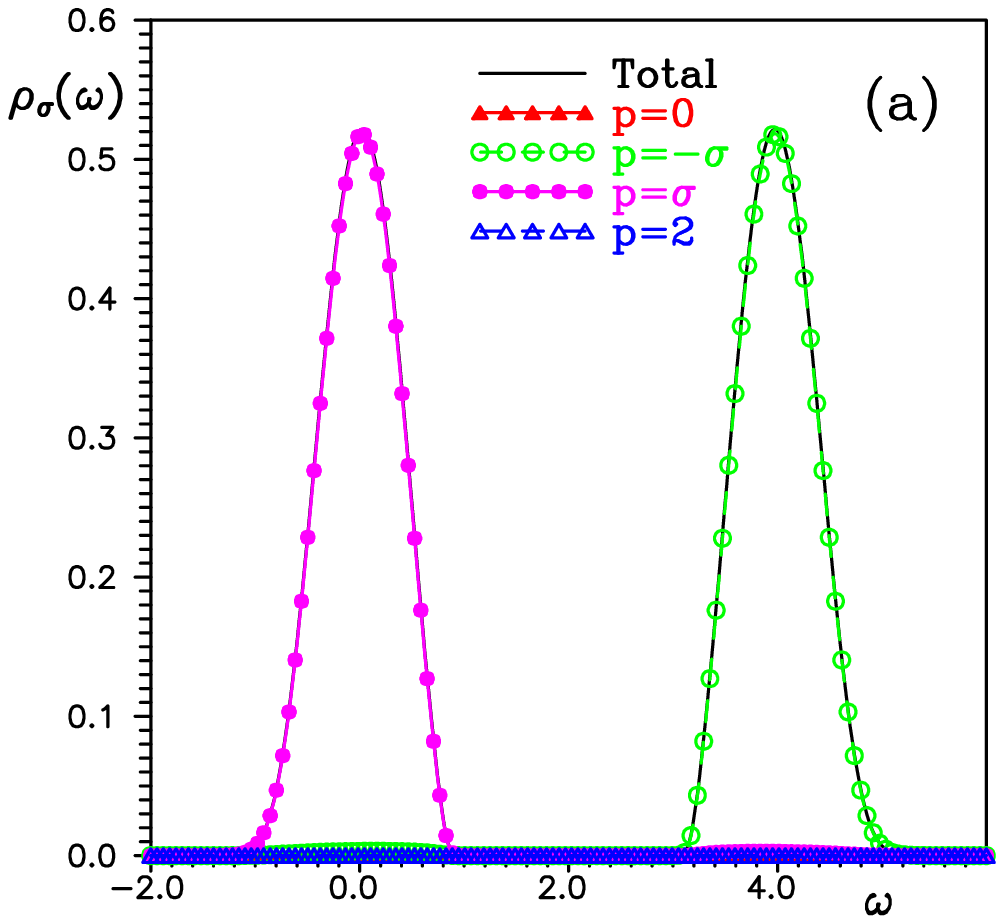}\hfill\null
\null\hfill\includegraphics[width=60mm]{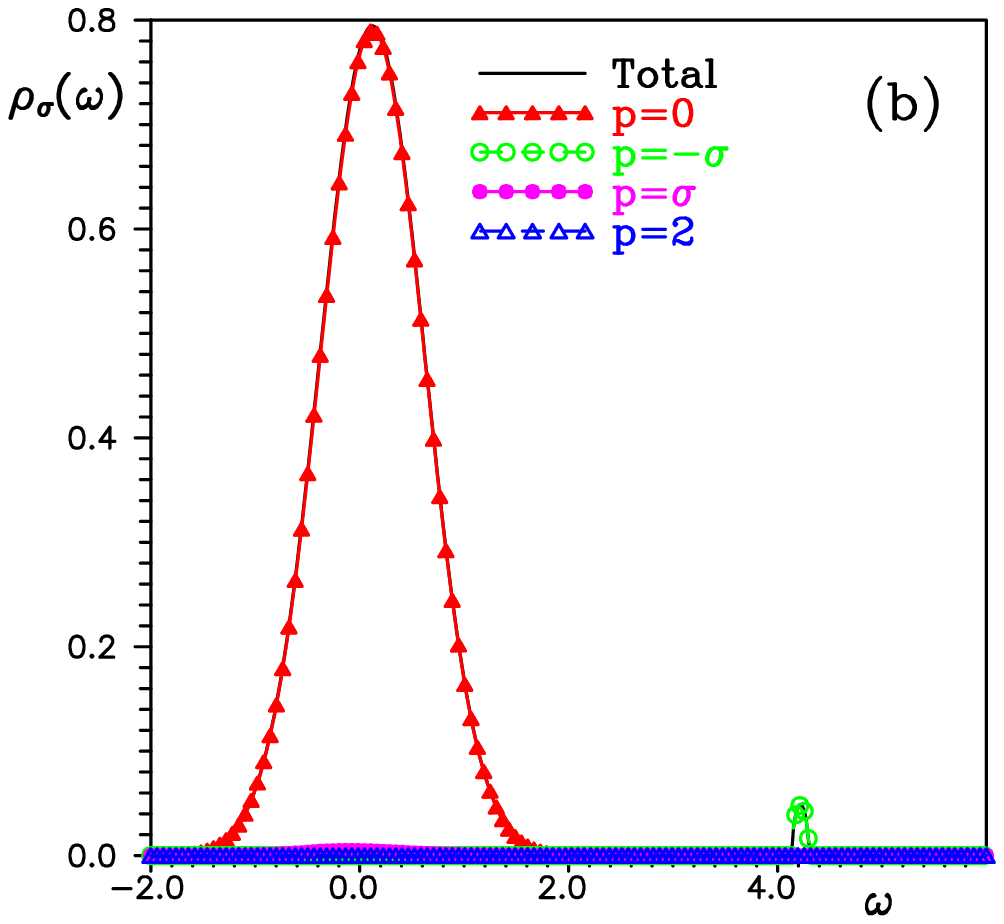}\hfill\null\\ [1em]
\null\hfill\includegraphics[width=60mm]{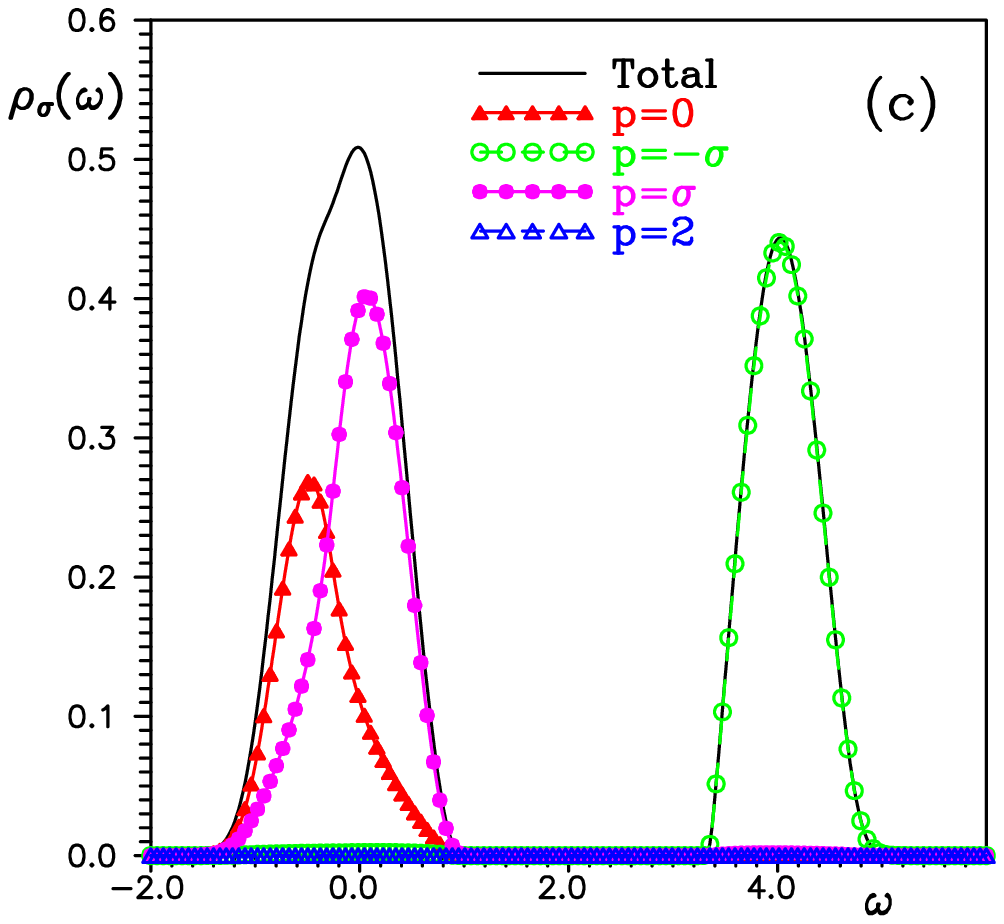}\hfill\null
\null\hfill\includegraphics[width=60mm]{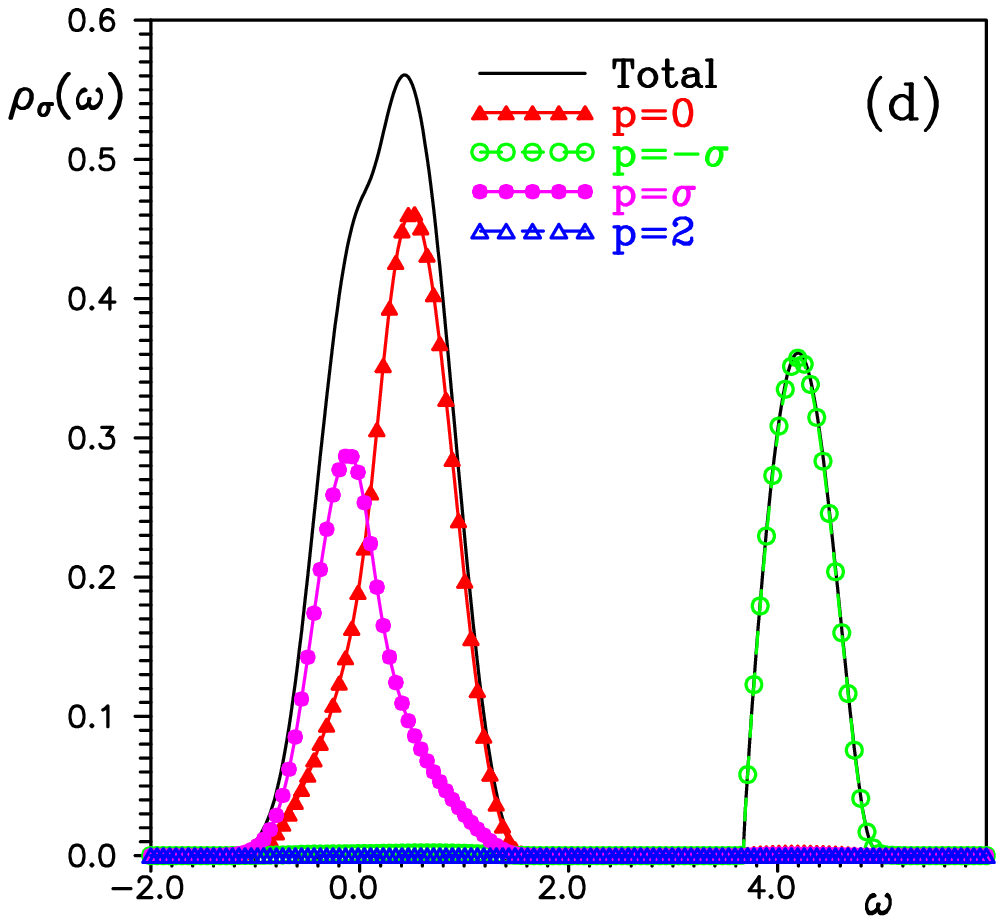}\hfill\null
\bigskip
\caption{Spectral weight function $\rho_{\sigma}(\omega)$
(\protect\ref{SWFt}): total and for each subspace, for the
different chemical potential values: (a) $\mu=\frac U2$, $n=1$;
(b) $\mu=-1$, $n=0.07$; (c) $\mu=0.01$, $n=0.72$;
(d) $\mu=-0.01$, $n=0.66$
($U=4$, $T=0.2$).}
\label{ImG_f}
\end{figure}

\begin{figure}
\noindent\null\hfill\includegraphics[width=60mm]{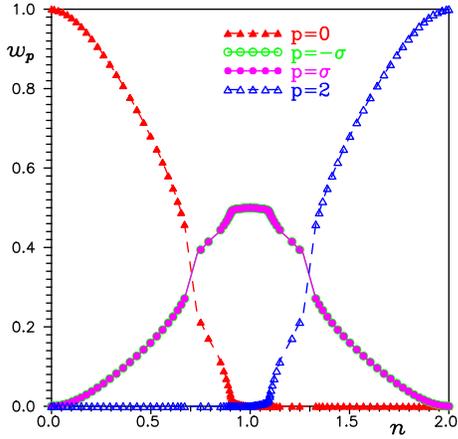}\hfill\null
\bigskip
\caption{Statistical weights of the subspaces $w_p$
(\protect\ref{eq38}) as functions of the electron concentration
($U=4$, $T=0.2$).}
\label{wp_n}
\end{figure}

In Fig.~\ref{ImG_f} the frequency distribution of the total
spectral weight function
\begin{equation}\label{SWFt}
\rho_{\sigma}(\omega)=\frac1{\pi}\Im G^{(a)}_{\sigma}(\omega-i0^+)
\end{equation}
as well as contributions into it from the
subspaces [separate terms in (\ref{GFHFt})] are presented for the
different electron concentration (chemical potential) values. One
can see, that the spectral weight function contains two peaks, which
correspond to the two Hubbard bands. Each band is formed by the
two close peaks: $p=0$ and $\sigma$ for the lower Hubbard band and
$p=2$ and $\bar\sigma$ for the upper one, with weights $w_p$
Eq.~(\ref{eq38}). The main contributions come (see Fig.~\ref{wp_n})
from the subspaces $p=0$ for the low electron concentrations
($n<\frac23$, $\mu<0$), $p=2$ for the low hole concentrations
($2-n<\frac23$, $\mu>U$) and $p=\sigma,\bar\sigma$ for the
intermediate values. For the small electron or hole
concentrations, the Green's function for the atomic problem
(\ref{GFHFt}) possesses correct Hartree-Fock limits too.

\end{multicols}
\widetext

\begin{multicols}{2}
\narrowtext

Such four-pole structure of the single-electron Green's function can
be obtained also for the one-dimensional chain with the $N=2$
periodic boundary condition (see the Appendix), which is equivalent to
the two-site problem considered by Harris and Lange.\cite{H_L}
Here, two poles
correspond to the noninteracting electrons or holes, which hope
over the empty sites, and
give the main contribution for small concentrations. The other two
poles give the main contribution close to half-filling and
correspond to the hopping of the strongly-correlated electrons
over the resonating valence bond (RVB) states.

So, one can suppose that the Hubbard model describes
strongly-correlated electronic systems
that contain four components (subspaces). Subspaces $p=0$ and
$p=2$ describe the Fermi-liquid component (electron and hole,
respectively) which is dominant for the small electron and hole
concentrations, when the chemical potential is close to the bottom of
the lower band and top of the upper one.
On the other hand,
subspaces $p=\uparrow$ and $\downarrow$ describe
the non-Fermi-liquid (strongly correlated, e.g., RVB) component,
which is dominant close to half-filling.
Within
the considered Hartree-Fock approximation, at $n\approx\frac23$
and $2-n\approx\frac23$, we have transition between these two
regimes: Fermi liquid and non-Fermi liquid. It reminds us the known
properties of the high-$T_c$ compounds, where for the nondoped
case ($n=1$) compounds are in the antiferroelectric dielectric
state, then for small doping the non-Fermi-liquid behavior is
observed (underdoped case $n\lesssim1$) and after some optimal doping value, the
properties of the compound sharply change from the non-Fermi to
the Fermi liquid (overdoped case).

The results presented in Figs.~\ref{ImG_f} and \ref{wp_n} are
obtained for relatively high temperature. With the
temperature decrease, on the one hand, the transition between the
Fermi and non-Fermi liquid becomes sharp and, on the other hand,
for some chemical potential values there can be three
solutions of Eq.~(\ref{mvHF}) with two of them corresponding to the phase-separated
states. The consideration of the phase separation in the Hubbard model
is not a topic of this paper and will be the subject of
further investigations.

\subsection{Beyond the Hartree-Fock approximation}

Self-energy in the Hartree-Fock approximation [see Eq.~(\ref{GFHF})]
describes some self-consistent shift of the initial energy levels
and does not depend on the frequency. All other improvements of
the expression for self-energy add the frequency dependent
contributions. To see this, let us consider the contribution into the
mean values from the first diagram in Eq.~(\ref{dPhi}). This diagram
originates from the following skeletal diagram
\begin{equation}\label{skelet}
\unitlength.2em
\begin{fmffile}{skelet}
\begin{fmfgraph}(15,10)\fmfstraight
    \fmfpen{thin}\fmfleft{l1,l2,l3}\fmfright{r1,r2,r3}
    \fmf{dbl_plain,left=.8,tension=.5}{l2,r2}
    \fmf{dbl_plain,right=.8,tension=.5}{l2,r2}
    \fmf{dashes}{l2,r2}\fmfdot{l2,r2}
\end{fmfgraph}
\end{fmffile}
\end{equation}
in the diagrammatic expansion for functional $\Phi_{(p)}$. On the
other hand, such a skeletal diagram produces additional contribution
into the self-energy
\begin{equation}\label{SigmaF}
\unitlength.2em
\begin{fmffile}{sigmaf}
\begin{fmfgraph}(15,10)\fmfstraight
    \fmfpen{thin}\fmfbottom{b1,b2}
    \fmf{dashes,left=1,tension=.5}{b1,b2}
    \fmf{dbl_plain}{b1,b2}\fmfdot{b1,b2}
\end{fmfgraph}
\end{fmffile}
\end{equation}
which is frequency dependent. Also, in order to get a
self-consistent set of equations, we introduce renormalized bosonic Green's
functions
\begin{eqnarray}
D_{20}(\omega_m)&=& \frac1{i\omega_m-\tilde\lambda_{20}};\;
D_{\sigma\bar\sigma}(\omega_m)=
\frac1{i\omega_m-\tilde\lambda_{\sigma\bar\sigma}},
\nonumber\\ \label{D02}
\tilde\lambda_{20}&=&\lambda_{20}+U\frac1{\beta}\sum_{n\sigma}
\Psi_{\sigma(p)}(\omega_{n}),
\\ \nonumber
\tilde\lambda_{\sigma\bar\sigma}&=&\lambda_{\sigma\bar\sigma}-
U\frac1{\beta}\sum_{n\sigma}\sigma\Psi_{\sigma(p)}(\omega_{n}).
\end{eqnarray}

Finally, for the Green's function
(\ref{GFp}) we get the general representation
\begin{equation}\label{GFpf}
G_{\sigma(p)}(\omega_n)=\frac1{i\omega_n+\mu_{\sigma}-Un_{\bar\sigma(p)}
-\widetilde\Sigma_{\sigma(p)}(\omega_n)-J_{\sigma}(\omega_n)},
\end{equation}
where the Hartree-Fock contribution $Un_{\bar\sigma(p)}$ is
extracted and $\widetilde\Sigma_{\sigma(p)}(\omega_n)$ is a
frequency dependent part of the self-energy, which within the considered
approximation is equal
\begin{equation}
\widetilde\Sigma_{\sigma(p)}(\omega_n)=US_{\sigma(p)}(\omega_n),
\end{equation}
where
\begin{equation}\label{SE02}
S_{\sigma(p)}(\omega_n)=
\pm\frac{U}{\beta}\sum_{n'} D_{\sigma\bar\sigma(p)}(\omega_{n},\omega_{n'})
\Psi_{\bar\sigma(p)}(\omega_{n'})
\end{equation}
and
\begin{equation}
D_{\sigma\bar\sigma(p)}(\omega_{n},\omega_{n'})=\left\{\begin{array}{cc}
  D_{20}(\omega_{n+n'}) & \text{for } p=0,2 \\
  D_{\sigma\bar\sigma}(\omega_{n-n'}) & \text{for }
  p=\sigma,\bar\sigma
\end{array}\right..
\end{equation}
Now, mean values (\ref{mv}) are equal
\begin{eqnarray}
n_{\sigma(p)} &=& n^{(0)}_{\sigma(p)} + \frac{1}{\beta} \sum\limits_{n}
\Psi_{\sigma(p)}(\omega_{n})
\label{eq49}\\
&&\pm\frac{1}{\beta^{2}} \sum\limits_{nn'}
U^2 D^2_{\sigma\bar\sigma(p)}(\omega_{n},\omega_{n'})
\Psi_{\sigma(p)} (\omega_{n})
\Psi_{\bar\sigma(p)} (\omega_{n'})
\nonumber
\end{eqnarray}
and for the grand canonical potentials in the subspaces we obtain
the following expressions:
\begin{eqnarray}\label{Omega02}
\Omega_{(p)}&=&\lambda_p-\frac1{\beta}\sum_{n\sigma}
\ln\left[1-J_{\sigma}(\omega_n)\Xi_{\sigma(p)}(\omega_n)\right]
\\ \nonumber
&&-\frac1{\beta^2}\sum_{nn'}U
\left[1\pm UD_{20}^2(\omega_{n+n'})g_{20}^{-1}(\omega_{n+n'})\right]
\\ \nonumber
&&\qquad\qquad\times\Psi_{\sigma(p)}(\omega_{n}) \Psi_{\bar\sigma(p)}(\omega_{n'})
\end{eqnarray}
for $p=0,2$, and
\begin{eqnarray}\label{OmegaUD}
\Omega_{(p)}&=&\lambda_p-\frac1{\beta}\sum_{n\sigma}
\ln\left(1-J_{\sigma}(\omega_n)\Xi_{\sigma(p)}(\omega_n)\right)
\\ \nonumber
&&-\frac1{\beta^2}\sum_{nn'}U
\left(1\pm UD_{\sigma\bar\sigma}^2(\omega_{n-n'})
g_{\sigma\bar\sigma}^{-1}(\omega_{n-n'})\right)
\\ \nonumber
&&\qquad\qquad\times\Psi_{\sigma(p)}(\omega_{n}) \Psi_{\bar\sigma(p)}(\omega_{n'})
\end{eqnarray}
for $p=\sigma,\bar\sigma$.

In order to analyze the structure of the poles in Eq.~(\ref{GFpf}),
an analytical continuation of the expression for
$\widetilde\Sigma_{\sigma(p)}(\omega_n)$ from the imaginary axis to the real one
should be done. To do it, we use the well-known identity
\begin{equation}
\frac1{\beta}\sum_n\frac{e^{i\omega_n 0^+}}{i\omega_n-\lambda}=
\pm n_\pm(\lambda),
\end{equation}
which follows from Eq.~(\ref{eq15}), and analytical properties of
the Green's function
\begin{equation}
G_\sigma(z)=\frac1{\pi}\int\limits_{-\infty}^{+\infty}\!\! d\omega
\frac{\Im G_\sigma(\omega-i0^+)}{z-\omega}.
\end{equation}
Green's functions in the subspaces $G_{\sigma(p)}(z)$, irreducible
parts $\Xi_{\sigma(p)}(z)$, and dynamical mean-field
$J_{\sigma}(z)$ all possess the same analytical properties.
Finally, for $S_{\sigma(p)}(z)$
we get the following expressions:
\begin{eqnarray}
S_{\sigma(p)}(z)&=&\pm \frac{U}{\pi}\;{\cal P}\!\!
\int\limits_{-\infty}^{+\infty}\!\!
d\omega\;
n_+(\omega)\;\frac{\Im\Psi_{\bar\sigma(p)}(\omega-i0^+)}{z+\omega-\tilde\lambda_{20}}
\nonumber\\
&&\pm n_-(\tilde\lambda_{20})\;U
\Psi_{\bar\sigma(p)}(\tilde\lambda_{20}-z)
\label{Sigma02}
\end{eqnarray}
for subspaces $p=0,2$ and
\begin{eqnarray}
S_{\sigma(p)}(z)&=&\pm \frac{U}{\pi}\;{\cal P}\!\!
\int\limits_{-\infty}^{+\infty}\!\!
d\omega\;
n_+(\omega)\;\frac{\Im\Psi_{\bar\sigma(p)}(\omega-i0^+)}{z-\omega-\tilde\lambda_{\sigma\bar\sigma}}
\nonumber\\
&&{\mp}\left[n_-(\tilde\lambda_{\sigma\bar\sigma})+1\right]\,U
\Psi_{\bar\sigma(p)}(z-\tilde\lambda_{\sigma\bar\sigma})
\label{SigmaUD}
\end{eqnarray}
for $p=\sigma,\bar\sigma$. Analytical continuation of
expressions (\ref{eq49}), (\ref{Omega02}), and (\ref{OmegaUD}) can be
done in the same way. One can see, that contributions
(\ref{Sigma02}) and (\ref{SigmaUD}) diverge when
$\tilde\lambda_{20}=0$ and $\tilde\lambda_{\sigma\bar\sigma}=0$,
respectively, which is an unphysical result.

So, we cannot include into the consideration only one contribution
from diagram (\ref{SigmaF}) but must sum up all diagrams of the
following type
\begin{eqnarray}\label{SigmaFS}
\unitlength.1em
&&\begin{fmffile}{sigmaFS}
\begin{fmfgraph}(30,10)\fmfstraight
    \fmfpen{thin}\fmfbottom{b1,b2}
    \fmf{dashes,left=1}{b1,b2}
    \fmf{dbl_plain}{b1,b2}\fmfdot{b1,b2}
\end{fmfgraph}
\;-2\;\;
\begin{fmfgraph}(30,10)\fmfstraight
    \fmfpen{thin}\fmfbottom{b1,b2}
    \fmf{dbl_plain}{b1,c1,c2,b2}
    \fmffreeze
    \fmf{dashes,left}{c1,b1,b2}
    \fmfdot{c1,b1,b2}
\end{fmfgraph}
\;+\;
\begin{fmfgraph}(30,10)\fmfstraight
    \fmfpen{thin}\fmfbottom{b1,b2}
    \fmf{dbl_plain}{b1,c1,c2,b2}
    \fmffreeze
    \fmf{dashes,left=1}{c1,b1,b2,c2}
    \fmfdot{c1,b1,b2,c2}
\end{fmfgraph}
\end{fmffile}
\\ \nonumber
\\ \nonumber
&&-\;
\unitlength.1em
\begin{fmffile}{sigmaFS1}
\begin{fmfgraph}(60,10)\fmfstraight
    \fmfpen{thin}\fmfbottom{b1,b2,b3}
    \fmf{dbl_plain}{b1,b2,b3}
    \fmffreeze
    \fmf{dashes,left=1}{b1,b2,b3}
    \fmfdot{b1,b2,b3}
\end{fmfgraph}
\;+\;
\begin{fmfgraph}(60,10)\fmfstraight
    \fmfpen{thin}\fmfbottom{b1,b2,b3}
    \fmf{dbl_plain}{b1,c1,c2,b2,b3}
    \fmffreeze
    \fmf{dashes,left=1}{c1,b1,b2,b3}
    \fmfdot{c1,b1,b2,b3}
\end{fmfgraph}
\;+\ldots
\end{fmffile}
\end{eqnarray}
which gives an expression free from the above-mentioned divergences
\begin{equation}
\widetilde\Sigma_{\sigma(p)} (\omega_{n})=
U\left(1+S^{\prime\prime}_{\sigma(p)} (\omega_{n})\right)
-U\frac{\left(1+S^{\prime}_{\sigma(p)} (\omega_{n})\right)^2}
{1+S_{\sigma(p)} (\omega_{n})},
\end{equation}
where $S_{\sigma(p)} (\omega_{n})$ is defined above and
\begin{eqnarray}
S^{\prime}_{\sigma(p)} (\omega_{n})&=&\frac{U}{\beta}\sum_{n'}
D_{\sigma\bar\sigma(p)}(\omega_{n},\omega_{n'})
\Psi_{\bar\sigma(p)}(\omega_{n'})
\nonumber\\
&&\qquad\quad\times
\frac{S_{\bar\sigma(p)}(\omega_{n'})-S^{\prime}_{\bar\sigma(p)}(\omega_{n'})}
{1+S_{\bar\sigma(p)} (\omega_{n'})},
\\
S^{\prime\prime}_{\sigma(p)} (\omega_{n})&=&\frac{U}{\beta}\sum_{n'}
D_{\sigma\bar\sigma(p)}(\omega_{n},\omega_{n'})
\Psi_{\bar\sigma(p)}(\omega_{n'})
\nonumber\\
&&\qquad\quad\times
\left(\frac{S_{\bar\sigma(p)}(\omega_{n'})-S^{\prime}_{\bar\sigma(p)}(\omega_{n'})}
{1+S_{\bar\sigma(p)} (\omega_{n'})}\right)^2.
\nonumber
\end{eqnarray}
Such diagram resummation must be also done in the expression for the
mean values (\ref{eq49}), where the last term must be replaced by
\begin{eqnarray}
&&\pm\frac{1}{\beta^{2}} \sum\limits_{nn'}
U^2 D^2_{\sigma\bar\sigma(p)}(\omega_{n},\omega_{n'})
\Psi_{\sigma(p)} (\omega_{n})
\Psi_{\bar\sigma(p)} (\omega_{n'})
\\ \nonumber
&&\qquad\qquad\times\left(\frac{1+S^{\prime}_{\sigma(p)}(\omega_n)}
{1+S_{\sigma(p)}(\omega_n)}+
\frac{1+S^{\prime}_{\bar\sigma(p)}(\omega_{n'})}
{1+S_{\bar\sigma(p)}(\omega_{n'})}-1\right)^2.
\end{eqnarray}

Besides diagram (\ref{SigmaF}), there are a lot of other diagrams
that diverge and need additional resummation of
the diagrammatic series. But now it is difficult to clear out what
types of diagrams are leading in different case, which calls for
additional investigation. But it is, obviously, that such
contributions will shift the boundary between the Fermi and
non-Fermi-liquid behavior.

\section{Summary}

A finite-temperature perturbation theory scheme in terms of electron hopping,
which is based on the Wick theorem for Hubbard operators and is valid for
arbitrary values of $U$ ($U<\infty$) has been developed for Hubbard-type models.
Diagrammatic series contain single-site vertices, which are
irreducible many-particle Green's functions for unperturbated
single-site Hamiltonian, connected by hopping lines.
Applying the Wick theorem for Hubbard operators has allowed us to calculate
these vertices and it is shown that for each vertex the problem splits
into subspaces with
``vacuum states'' determined by the diagonal (projection)
operators and only excitations around these ``vacuum states'' are
allowed. The vertices possess a finite $U\to\infty$ limit when diagrammatic series
of the strong-coupling approach\cite{IzyumovLetfulov,IzyumovPRB} are reproduced.
The rules to construct diagrams by the primitive vertices are
proposed.

In the limit of infinite spatial dimensions the total auxiliary
single-site problem exactly (naturally) splits into subspaces
(four for Hubbard model) and a considered analytical
scheme allows to build a self-consistent Kadanoff-Baym-type theory
for the Hubbard model. Some analytical results are given for simple
approximations: an alloy-analogy approximation,
when two-pole structure for Green's function is obtained,
which is exact for the Falicov-Kimball model, and the Hartree-Fock-type
approximation, which results in the four-pole
structure for the Green's function. Expanding beyond the Hartree-Fock
approximation calls for considering of frequency
dependent contributions into the self-energy and resummation of
the diagrammatic series.

In general, the expression
\begin{eqnarray}
&&G^{(a)}_{\sigma}(\omega_n)
\\ \nonumber
&&\quad=\sum_p
\frac{w_p}{i\omega_n+\mu_{\sigma}-Un_{\bar\sigma(p)}
-\widetilde\Sigma_{\sigma(p)}(\omega_n)-J_{\sigma}(\omega_n)}
\end{eqnarray}
gives an exact
four-pole structure for the single-electron Green's function of the
effective atomic problem. In Eq.~(\ref{eq40}) zero-order Green's
functions (\ref{eq23}) are the same for the subspaces $p=0,\sigma$ and
$p=2,\bar\sigma$, respectively, and correspond to the two-pole
solution of the one-site problem without hopping. Switching on of
the electron hopping splits these two poles and the value of
splitting is determined by the values of the self-energy parts in
the subspaces, which describe the
contributions from the different scattering processes.
Alloy-analogy approximation neglects by the such scattering
processes ($\Sigma_{\sigma(p)}(\omega_n)=0$) which results in the
two-pole structure for the Green's functions (\ref{AAA}). But, in
general, Green's functions possess four-pole structure and even the Hartree-Fock
approximation (\ref{GFHFt}) clearly shows it.

It should be noted that the four-pole structure of the Green's function
for the atomic problem might not result in the four bands of the
spectral weight function (see Fig.~\ref{ImG_f}). The presented
consideration allows us to suppose that each pole describes
contributions from the different components (subspaces) of the
electronic system: Fermi liquid (subspaces $p=0,2$) and non-Fermi
liquid ($p=\uparrow,\downarrow$), and for
small electron and hole concentrations ($n<\frac23$ and $2-n<\frac23$)
the Fermi-liquid component gives the main contribution (``overdoped
regime'' of high-$T_c$'s), whereas in other cases the non-Fermi liquid one
(``underdoped regime'').

\acknowledgements

I am grateful to Professor I.V.~Stasyuk for the very useful and
stimulating discussions.
Also I acknowledge the hospitality of the
Institute of Physics of the M.~Curie Sk\l odowska University of
Lublin and thank Professor K.I.~Wysoki\'nski for discussions.

\appendix

\section*{Two-site problem}

Let us consider an infinite one-dimensional chain with the $N=2$
periodic boundary condition. Mathematically it is equivalent to
the two-site problem considered by Harris and Lange,\cite{H_L} but
now we can introduce the lattice Fourier transformation, with two
wave-vector values in the first Brillouin zone $q=0$ and $q=\pi$, and
perform all calculations for the grand canonical ensemble. The
Hamiltonian of the model is the following:
\begin{eqnarray}\label{H2s}
H&=&\sum_{i=1,2}\left(Un_{i\uparrow}n_{i\downarrow}-
\mu\sum_{\sigma}n_{i\sigma}\right)
\\
&&+t\sum_{\sigma}\left(a_{1\sigma}^{\dag}a_{2\sigma}+
a_{2\sigma}^{\dag}a_{1\sigma}\right).
\nonumber
\end{eqnarray}
We can introduce the Fourier transform of the electron hopping
\begin{equation}
t_q=t\cos q=\left\{\begin{array}{cc}
  t & \text{ for } q=0 \\
  -t & \text{ for } q=\pi
\end{array}\right.
\end{equation}
and our aim is to calculate the single-electron Green's function
\begin{equation}\label{GF2s}
G_{\sigma}(\omega,t_q)=\left\{\begin{array}{cc}
  G_{11\sigma}(\omega_n)+G_{12\sigma}(\omega_n) & \text{ for } q=0 \\
  G_{11\sigma}(\omega_n)-G_{12\sigma}(\omega_n) & \text{ for } q=\pi
\end{array}\right.,
\end{equation}
where $G_{ij\sigma}(\tau-\tau')=
\left\langle Ta_{i\sigma}^{\dag}(\tau)a_{j\sigma}(\tau')
\right\rangle$.

The initial basis of states contains 16 many-electron two-site
states $\left|p_1,p_2\right\rangle$, where
$p_i=\left\{n_{i\uparrow}n_{i\downarrow}\right\}$, i.e.,
\begin{eqnarray}
|1\rangle&=&|0,0\rangle,
\nonumber\\
|2\rangle&=&|\downarrow,0\rangle=a_{1\downarrow}^{\dag}|1\rangle.
\nonumber\\
|3\rangle&=&|0,\downarrow\rangle=a_{2\downarrow}^{\dag}|1\rangle,
\nonumber\\
|4\rangle&=&|\uparrow,0\rangle=a_{1\uparrow}^{\dag}|1\rangle,
\nonumber\\
|5\rangle&=&|0,\uparrow\rangle=a_{2\uparrow}^{\dag}|1\rangle,
\nonumber\\
|6\rangle&=&|\downarrow,\downarrow\rangle=a_{2\downarrow}^{\dag}|2\rangle=-a_{1\downarrow}^{\dag}|3\rangle,
\nonumber\\
|7\rangle&=&|\uparrow,\uparrow\rangle=a_{2\uparrow}^{\dag}|4\rangle=-a_{1\uparrow}^{\dag}|5\rangle,
\nonumber\\
|8\rangle&=&|2,0\rangle=a_{1\uparrow}^{\dag}|2\rangle=-a_{1\downarrow}^{\dag}|4\rangle,
\nonumber\\
|9\rangle&=&|\uparrow,\downarrow\rangle=a_{1\uparrow}^{\dag}|3\rangle=-a_{2\downarrow}^{\dag}|4\rangle,
\\
|10\rangle&=&|\downarrow,\uparrow\rangle=a_{2\uparrow}^{\dag}|2\rangle=-a_{1\downarrow}^{\dag}|5\rangle,
\nonumber\\
|11\rangle&=&|0,2\rangle=a_{2\uparrow}^{\dag}|3\rangle=-a_{2\downarrow}^{\dag}|5\rangle,
\nonumber\\
|12\rangle&=&|2,\downarrow\rangle=a_{1\uparrow}^{\dag}|6\rangle=a_{1\downarrow}^{\dag}|9\rangle=-a_{2\downarrow}^{\dag}|8\rangle,
\nonumber\\
|13\rangle&=&|\downarrow,2\rangle=a_{2\uparrow}^{\dag}|6\rangle=a_{1\downarrow}^{\dag}|11\rangle=-a_{2\downarrow}^{\dag}|10\rangle,
\nonumber\\
|14\rangle&=&|2,\uparrow\rangle=a_{1\downarrow}^{\dag}|7\rangle=a_{2\uparrow}^{\dag}|8\rangle=-a_{1\uparrow}^{\dag}|10\rangle,
\nonumber\\
|15\rangle&=&|\uparrow,2\rangle=a_{2\downarrow}^{\dag}|7\rangle=a_{2\uparrow}^{\dag}|9\rangle=-a_{1\uparrow}^{\dag}|11\rangle,
\nonumber\\
|16\rangle&=&|2,2\rangle=a_{1\uparrow}^{\dag}|13\rangle=-a_{2\uparrow}^{\dag}|12\rangle=a_{1\downarrow}^{\dag}|15\rangle=a_{2\downarrow}^{\dag}|14\rangle
\nonumber
\end{eqnarray}
and one can introduce Hubbard operators $X^{p,q}=|p\rangle\langle q|$
acting in the space of these states. Now, electron creation operators
can be presented in the following form:
\begin{eqnarray}
a_{1\uparrow}^{\dag}&=&X^{4,1}-X^{7,5}+X^{8,2}+X^{9,3}
\nonumber\\
&&+X^{12,6}-X^{14,10}-X^{15,11}+X^{16,13},
\nonumber\\
a_{1\downarrow}^{\dag}&=&X^{2,1}-X^{6,3}-X^{8,4}-X^{10,5}
\nonumber\\
&&+X^{12,9}+X^{13,11}+X^{14,7}+X^{16,15},
\\
a_{2\uparrow}^{\dag}&=&X^{5,1}+X^{7,4}+X^{10,2}+X^{11,3}
\nonumber\\
&&+X^{13,6}+X^{14,8}+X^{15,9}-X^{16,12},
\nonumber\\
a_{2\downarrow}^{\dag}&=&X^{3,1}+X^{6,2}-X^{9,4}-X^{11,5}
\nonumber\\
&&-X^{12,8}-X^{13,10}+X^{15,7}-X^{16,14}.
\nonumber
\end{eqnarray}
By transformations
\begin{eqnarray}
\lefteqn{\left(\begin{array}{cccc}
  |2\rangle & |4\rangle & |12\rangle & |14\rangle \\
  |3\rangle & |5\rangle & |13\rangle & |15\rangle
\end{array}\right)}
\\ \nonumber
&&\qquad=\left\|\begin{array}{cc}
  \frac1{\sqrt{2}} & -\frac1{\sqrt{2}} \\
  \frac1{\sqrt{2}} & \frac1{\sqrt{2}}
\end{array}\right\|\left(\begin{array}{cccc}
  |\tilde2\rangle & |\tilde4\rangle & |\widetilde{12}\rangle & |\widetilde{14}\rangle \\
  |\tilde3\rangle & |\tilde5\rangle & |\widetilde{13}\rangle & |\widetilde{15}\rangle
\end{array}\right)
\end{eqnarray}
and
\begin{equation}
\left(\begin{array}{c}
  |8\rangle \\
  |9\rangle \\
  |10\rangle \\
  |11\rangle
\end{array}\right)=\left\|\begin{array}{cccc}
  \frac1{\sqrt{2}}\cos\phi & \frac{-1}{\sqrt{2}}\sin\phi & 0 & \frac{-1}{\sqrt{2}} \\
  \frac1{\sqrt{2}}\sin\phi & \frac1{\sqrt{2}}\cos\phi & \frac{-1}{\sqrt{2}} & 0 \\
  \frac1{\sqrt{2}}\sin\phi & \frac1{\sqrt{2}}\cos\phi & \frac1{\sqrt{2}} & 0 \\
  \frac1{\sqrt{2}}\cos\phi & \frac{-1}{\sqrt{2}}\sin\phi & 0 & \frac1{\sqrt{2}}
\end{array}\right\|\left(\begin{array}{c}
  |\tilde8\rangle \\
  |\tilde9\rangle \\
  |\widetilde{10}\rangle \\
  |\widetilde{11}\rangle
\end{array}\right),
\end{equation}
where $\sin2\phi(t)=\frac{2t}{\sqrt{U^2/4+4t^2}}$, the Hamiltonian
(\ref{H2s}) can be diagonalized
\begin{equation}
H=\sum_p \lambda_{\tilde p} X^{\tilde p, \tilde p}.
\end{equation}
Here
\begin{eqnarray}
\lambda_{\tilde1}&=&0,
\nonumber\\
\lambda_{\tilde2}=\lambda_{\tilde4}&=&-\mu+t,
\nonumber\\
\lambda_{\tilde3}=\lambda_{\tilde5}&=&-\mu-t,
\nonumber\\
\lambda_{\tilde6}=\lambda_{\tilde7}=\lambda_{\widetilde{10}}&=&-2\mu,
\nonumber\\
\lambda_{\tilde8}&=&U+J-2\mu,
\\
\lambda_{\tilde9}&=&-J-2\mu,
\nonumber\\
\lambda_{\widetilde{11}}&=&U-2\mu,
\nonumber\\
\lambda_{\widetilde{12}}=\lambda_{\widetilde{14}}&=&U-3\mu+t,
\nonumber\\
\lambda_{\widetilde{13}}=\lambda_{\widetilde{15}}&=&U-3\mu-t,
\nonumber\\
\lambda_{\widetilde{16}}&=&2U-4\mu
\nonumber
\end{eqnarray}
are eigenvalues and
\begin{equation}\label{Jint}
J=\frac{4t^2}{\sqrt{U^2/4+4t^2}+U/2}
\longrightarrow\frac{4t^2}U \quad (U\gg t).
\end{equation}
Finally, with the use of the Wick theorem (\ref{eq14}) for the Hubbard operators
acting in the space of eigenstates $|\tilde p\rangle$, for the
single-electron Green's function (\ref{GF2s}) we obtain
\begin{eqnarray}\label{GF2ss}
G_{\sigma}(\omega,t_q)&=&
\frac{A_1(t_q)}{\omega-t_q}+\frac{B_1(t_q)}{\omega+J+t_q}
\\ \nonumber
&&+\frac{A_2(t_q)}{\omega-U-t_q}+\frac{B_2(t_q)}{\omega-U-J+t_q},
\end{eqnarray}
where
\begin{eqnarray}
\lefteqn{A_1(t)=\frac1Z\left[\vphantom{\frac12}
1+e^{\beta(\mu-t)}+e^{\beta(\mu+t)}+e^{2\beta\mu}\right.}
\nonumber\\
&&\qquad\left.+\frac12\left(e^{\beta(\mu+t)}+e^{2\beta\mu}+e^{-\beta(U-2\mu)}+e^{-\beta(U-3\mu+t)}\right)\right],
\nonumber\\
\lefteqn{A_2(t)=\frac1Z\left[\vphantom{\frac12}
e^{-\beta(2U-4\mu)}+e^{-\beta(U-3\mu+t)}\right.}
\nonumber\\
&&\qquad\qquad\qquad+e^{-\beta(U-3\mu-t)}+e^{2\beta\mu}
\\
&&\qquad\left.+\frac12\left(e^{\beta(\mu+t)}+e^{2\beta\mu}+e^{-\beta(U-2\mu)}+e^{-\beta(U-3\mu+t)}\right)\right],
\nonumber\\
\lefteqn{B_{1,2}(t)=\frac1{2Z}\left(1\pm\frac{2t}{\sqrt{U^2/4+4t^2}}\right)}
\nonumber\\
&&\quad\times\left[e^{\beta(\mu-t)}+e^{\beta(2\mu+J)}+e^{-\beta(U-2\mu+J)}+e^{-\beta(U-3\mu-t)}\right]
\nonumber
\end{eqnarray}
and $Z=\sum_p e^{-\beta\lambda_{\tilde p}}$.

One can see, that
Green's function (\ref{GF2ss}) possesses a four-pole structure and the spectrum contains
four ``bands'' grouped near the initial energy levels of the
one-site problem $0$ and $U$. The distance between the centers of
gravity of the grouped bands is equal to $J$ [Eq.~(\ref{Jint})] and is of the order
of magnitude of the effective exchange interaction. It is obvious
that the weights of the bands satisfy the sum rule
\begin{equation}
A_1(t_q)+A_2(t_q)+B_1(t_q)+B_2(t_q)=1.
\end{equation}
The spectral weight function is equal
\begin{eqnarray}
\rho_{\sigma}(E)&=&\frac1{\pi N}\sum_q \Im G_{\sigma}(E-i0^+,t_q)
\nonumber\\
&=&\frac12\left[A_1(t)\,\delta(E-t)+A_1(-t)\,\delta(E+t)\right.
\label{SWF}\\
&&\quad+A_2(t)\,\delta(E-U-t)+A_2(-t)\,\delta(E-U+t)
\nonumber\\
&&\quad+B_1(t)\,\delta(E+J+t)+B_1(-t)\,\delta(E+J-t)
\nonumber\\
&&\quad+B_2(t)\,\delta(E-U-J+t)
\nonumber\\
&&\quad+\left.B_2(-t)\,\delta(E-U-J-t)\right],
\nonumber
\end{eqnarray}
contains the same eight energies obtained by Harris and
Lange\cite{H_L} but with different weights and originates from
the four poles (bands) of the Green's function (\ref{GF2ss}) for the two-site
problem.

The nature of these peaks is clear from the ground-state
properties of the model. At zero temperature, depending on the
value of the chemical potential or electron concentration, the
ground states are the following ($U\gg t$): empty state $n=0$ ($\mu<-t$):
\begin{eqnarray}
&& |\tilde1\rangle=|0,0\rangle,
\nonumber\\
&& \lambda_{\tilde1}=0,
\nonumber\\
&& G_{\sigma}(\omega,q)=\frac1{\omega-t_q},
\label{0GF}
\end{eqnarray}
one-electron states $n=\frac12$ ($-t<\mu<t-J$):
\begin{eqnarray}
&& |\tilde3\rangle=\frac1{\sqrt2}\left(a_{2\downarrow}^{\dag}-a_{1\downarrow}^{\dag}\right)|0,0\rangle,
\nonumber\\
&& |\tilde5\rangle=\frac1{\sqrt2}\left(a_{2\uparrow}^{\dag}-a_{1\uparrow}^{\dag}\right)|0,0\rangle,
\nonumber\\
&& \lambda_{\tilde3}=\lambda_{\tilde5}=-\mu-t,
\nonumber\\
&& G_{\sigma}(\omega,q=0)=\frac{3/4}{\omega-t}+\frac{1/4}{\omega-U-t},
\label{1GF}\\
&& G_{\sigma}(\omega,q=\pi)=\frac{1/2}{\omega+t}
+\frac{\frac14(1+\sin2\phi)}{\omega+J-t}+\frac{\frac14(1-\sin2\phi)}{\omega-U-J-t},
\nonumber
\end{eqnarray}
two-electron state $n=1$ ($t-J<\mu<U+J-t$):
\begin{eqnarray}
&& |\tilde9\rangle=\frac1{\sqrt2}\cos\phi
\left(a_{1\uparrow}^{\dag}a_{2\downarrow}^{\dag}
-a_{1\downarrow}^{\dag}a_{2\uparrow}^{\dag}\right)|0,0\rangle
\nonumber\\
&&\qquad-\frac1{\sqrt2}\sin\phi
\left(a_{1\uparrow}^{\dag}a_{1\downarrow}^{\dag}-a_{2\downarrow}^{\dag}a_{2\uparrow}^{\dag}\right)|0,0\rangle,
\nonumber\\
&& \lambda_{\tilde9}=-2\mu-J,
\nonumber\\
&& G_{\sigma}(\omega,q)=
\frac{\frac12(1+\sin2\phi_q)}{\omega+J+t_q}+\frac{\frac12(1-\sin2\phi_q)}{\omega-U-J+t_q},
\label{2GF}
\end{eqnarray}
three-electron states $n=\frac32$ ($U+J-t<\mu<U+t$):
\begin{eqnarray}
&& |\widetilde{13}\rangle=\frac1{\sqrt2}\left(a_{1\uparrow}+a_{2\uparrow}\right)|2,2\rangle,
\nonumber\\
&& |\widetilde{15}\rangle=\frac1{\sqrt2}\left(a_{1\downarrow}+a_{2\downarrow}\right)|2,2\rangle,
\nonumber\\
&& \lambda_{\widetilde{13}}=\lambda_{\widetilde{15}}=U-3\mu-t,
\nonumber\\
&& G_{\sigma}(\omega,q=0)=\frac{\frac14(1-\sin2\phi)}{\omega+J+t}
\label{3GF}\\
&&\qquad\qquad\qquad+\frac{\frac14(1+\sin2\phi)}{\omega-U-J+t}+\frac{1/2}{\omega-U-t},
\nonumber\\
&& G_{\sigma}(\omega,q=\pi)=\frac{1/4}{\omega+t}+\frac{3/4}{\omega-U+t},
\nonumber
\end{eqnarray}
and four-electron state $n=2$ ($\mu>U+t$):
\begin{eqnarray}
&& |\widetilde{16}\rangle=|2,2\rangle,
\nonumber\\
&& \lambda_{\widetilde{16}}=2U-4\mu,
\nonumber\\
&& G_{\sigma}(\omega,q)=\frac1{\omega-U-t_q}.
\label{4GF}
\end{eqnarray}

For small electron ($n\approx0$) or hole ($n\approx2$) concentrations
we get Green's functions (\ref{0GF}) and (\ref{4GF}), respectively,
which describe hopping of the noninteracting particles over empty
states.

On the other hand, for the
half-filled (symmetric) case $n\approx1$, the ground state
$|\tilde9\rangle$ is mainly a RVB-type state. Now, Green's
function (\ref{2GF})
possesses two-poles shifted by the value of the effective exchange
interaction $J$ from the one-site levels and describes the electron
transfer over the RVB states. The weight of each pole depends on
the hopping value, but its total contribution into the
spectral weight function (\ref{SWF}) is equal to $\frac12$ as it should be for
the symmetric case.

For other cases the number and weights of the poles in the spectral
weight function (\ref{SWF}) strongly
depend on the electron concentration (chemical potential) and
wave-vector values and contain contributions from the
noninteracting electrons (holes) and the strongly hybridized
RVB states.

\end{multicols}

\end{document}